\documentclass[a4paper]{article}

\usepackage[english]{babel}
\usepackage[utf8x]{inputenc}
\usepackage{amsmath}
\usepackage{amssymb}
\usepackage{amsfonts}
\usepackage{multirow}
\usepackage{multicol}
\usepackage{amsthm}
\usepackage{graphicx}
\usepackage{color}
\usepackage[colorinlistoftodos]{todonotes}
\usepackage{enumerate}
\usepackage{tikz,xcolor}
\usepackage{pgfplots}
\usepackage{gensymb}
\pgfplotsset{compat=1.14}
\usetikzlibrary{intersections,shapes.arrows,calc}
\tikzset{arw/.style={>={Triangle[length=1mm,width=1.5mm]},line width=1mm,draw=black!50}}
\tikzset{rect/.style={rectangle, [fill =blue!50, width = 1pt, height = 1pt]} }
\usetikzlibrary{3d,decorations.text,shapes.arrows,positioning,fit,backgrounds}
\tikzset{circle dotted/.style={dash pattern=on .05mm off 2mm,
                                         line cap=round}}
\usepackage{float}

\usetikzlibrary{positioning,arrows,arrows.meta}

\def\C{\mathbb{C}}
\def\R{\mathbb{R}}

\def\sone{\mathbb{S}^1}
\def\vp{\varphi}
\newcommand{\Pc}{\mathcal{P}}
\newcommand{\Hm}{\mathcal{H}}

\tikzstyle{sensor}=[draw, fill=blue!20, text width=2em, font = \small, 
    text centered, minimum height=2.5em]
\tikzstyle{II} = [sensor, text width=5em, font =\small, fill=blue!20, minimum height=3em, minimum width = 1.5em]
\tikzstyle{CBN} = [sensor, text width=5em, font = \small, fill=red!20, 
    minimum height=3em, rounded corners]
\tikzstyle{FC}=[draw, fill=black!20, text width=5em, font = \small, 
    text centered, minimum height=2.5em, minimum width = 10.5em,rounded corners]
    \tikzstyle{OP} = [sensor, text width=5em, font =\small, fill=blue!20, minimum height=3em, minimum width = 2em]
    
    \tikzstyle{FC_1}=[draw, fill=black!20, text width=5em, font = \small, 
    text centered, minimum height=10.5em, minimum width = 2em,rounded corners]
    \tikzstyle{OP} = [sensor, text width=5em, font =\small, fill=blue!20, minimum height=3em, minimum width = 2em]
\tikzstyle{CBN_1} = [sensor, text width=5em, font = \small, fill=red!20, 
    minimum height=3em, rounded corners]
    \tikzstyle{CBN_2} = [sensor, text width=5em, font = \small, fill=red!20, 
    minimum height=3em, rounded corners]

    \tikzstyle{maxpool} = [sensor, text width=5em, font = \small, fill=red!20, 
    minimum height=2.5em, rounded corners]
    \tikzstyle{maxpool_1} = [sensor, text width=5em, font = \small, fill=red!20, minimum height=2.5em, rounded corners]
    
   \tikzstyle{maxpool_2} = [sensor, text width=5em, font = \small, fill=red!20, minimum height=2em, rounded corners]

\title{Classification of stroke using Neural Networks in EIT}
\author{Juan Pablo Agnelli, Aynur \c{C}\"ol, Matti Lassas, Rashmi Murthy \\
Matteo Santacesaria and Samuli Siltanen}

\graphicspath{ {./images/} }

\begin{document}
\maketitle

\begin{abstract}
Electrical Impedance Tomography (EIT) is an emerging non-invasive medical imaging modality. It is based on feeding electrical currents into the patient, measuring the resulting voltages at the skin, and recovering the internal conductivity distribution. The mathematical task of EIT image reconstruction is a nonlinear and ill-posed inverse problem. Therefore any EIT image reconstruction method needs to be regularized, typically resulting in blurred images. One promising application is stroke-EIT, or classification of stroke into either ischemic or hemorrhagic. Ischemic stroke involves a blood clot, preventing blood flow to a part of the brain causing a low-conductivity region. Hemorrhagic stroke means bleeding in the brain causing a high-conductivity region. In both cases the symptoms are identical, so a cost-effective and portable classification device is needed. Typical EIT images are not optimal for stroke-EIT because of blurriness. This paper explores the possibilities of machine learning in improving the classification results. Two paradigms are compared: (a) learning from the EIT data, that is Dirichlet-to-Neumann (DN) maps and (b) extracting robust features from data and learning from them. The features of choice are Virtual Hybrid Edge Detection (VHED) functions [Greenleaf {\it et al.}, Analysis \& PDE 11, 2018] that have a geometric interpretation and whose computation from EIT data does not involve calculating a full image of the conductivity. We report  the measures of accuracy, sensitivity and specificity of the networks trained with EIT data and VHED functions separately. Computational evidence based on simulated noisy EIT data suggests that the regularized grey-box paradigm (b) leads to significantly better classification results than the black-box paradigm (a).

\end{abstract}


\section{Introduction}

The central question in inverse problems is how to extract information from the indirect measurements. In the interesting ill-posed cases the desired information is buried in the data in a nonlinear and unstable manner. The classical way is to design a {\it regularization strategy} whose implementation provides a noise-robust recovery algorithm. In recent years, {\it machine learning} has arisen as a data-driven alternative; it uses a set of examples for building a computational model to retrieve the desired information. Both approaches have their strengths and weaknesses. We study a possibility of getting the best of both worlds in the nonlinear and severely ill-posed case of electrical impedance tomography (EIT). 

Medical EIT imaging is based on feeding harmless electric currents into the patient through electrodes placed on the skin, and measuring the resulting voltages at the electrodes. The goal  is to recover an image of the electrical conductivity inside the patient. As different tissues and organs have different conductivities, the image contains medically relevant information. 

EIT can be mathematically modelled by the inverse conductivity problem \cite{Calder'on1980}. In this paper we restrict to the two-dimensional case; while patients are three-dimensional, often the 2D approximation gives useful results (see \cite{Alsaker19,Alsaker18,edic1998iterative,Isaacson2006}). Consider a simply connected domain $\Omega \subset \mathbb{R}^2 $, with a voltage distribution $f$ at the boundary. Let the scalar conductivity $\sigma \in {L}^{\infty}(\Omega)$ be bounded away from zero: $\sigma(x) \geq c > 0$. Then the EIT problem can be modeled by generalized Laplace equation
\begin{equation}\label{eq:cond}
    \nabla \cdot \sigma \nabla u = 0 \quad \text{in } \Omega, \quad \quad u|_{\partial \Omega} = f.
\end{equation}
where $u$ is the electric potential. 
Infinite precision voltage-to-current boundary measurements are modeled by the Dirichlet-to-Neumann (DN) map 
\begin{equation}\label{DN_operator}
    \Lambda_{\sigma} : f \rightarrow \sigma \frac{\partial u}{\partial \nu}
\end{equation}
where $\nu$ is the unit outward normal vector of $\partial \Omega$. 

We note that a more precise model for the practical current-to-voltage measurements is given by the complete electrode model \cite{Hyvonen2004,Somersalo1992}. However, in this {\it{proof-of-concept}} work we stay within the framework of continuum voltage-to-current models based on (\ref{DN_operator}).

We choose as our model problem the classification of stroke into hemorrhagic (bleed in the brain) or ischemic (blood clot in a vessel in the brain). The treatment protocols for the brain hemorrhage and ischemia are very different. Ischemic stroke can be treated, for example with thrombolysis (intravenous blood clot dissolving agent) or thrombectomy (intra-arterial mechanical removal of blood clot), often reinstalling the normal brain perfusion. Doing this quickly after onset can spare the patient from brain damage or even death \cite{saver2006time}. The treatment protocol for the brain hemorrhage involves stopping the excessive bleeding in the brain \cite{cite-key}.
Thus, the classification of the two different types of stroke is very important.

EIT has shown great potential to discriminate between the two different types of stroke \cite{mcewan2006design,shi2009experimental,bayford2012bioimpedance,malone2014stroke,boverman2016detection}. This is based on the conductivity differences between various healthy tissues and stroke-affected areas inside the patient's head. One can form an EIT reconstruction using a noise-robust regularization strategy such as iterative methods \cite{zhou2015comparison} or the direct D-bar method based on a nonlinear low-pass filter \cite{knudsen2009regularized,mueller2012linear}. However, the resistive skull makes it particularly difficult to produce a reliable EIT image of the brain. The resisitive skull reduces the probing energy of electric currents actually reaching the brain. Imaging the brain through the skull is one of the most challenging tasks for EIT and it is not covered in the present work, which instead focuses only on the problem of classification of strokes.

In recent years, machine learning techniques have proven to be very successful in solving inverse problems \cite{arridge2019,8103129,8253590} and they have been already applied in EIT imaging \cite{hamilton2019beltrami,seo2019learning}. Concerning the problem of classification of strokes, we are only aware of the work \cite{mcdermott2018brain} in which a SVM classifier is trained directly on electrode data.

In this paper, we suggest a hybrid approach for the classification of stroke, called {\it Robust Grey-Box (RGB) inversion}. We first apply Virtual Hybrid Edge Detection (VHED), introduced recently in \cite{greenleaf2018propagation}, for extracting noise-robust geometric features from the DN map. The VHED profiles pick out specific oriented  information about the conductivity, without necessarily forming an image. Then we classify stroke by applying machine learning to these noise-robust geometric features. This way we can maximize both noise-robustness and interpretability in the recovery of information, while taking advantage of the data adaptability and computational power of neural network models. 

Throughout our numerical tests, we use the continuum model for EIT measurements. This is unrealistic from the practical point of view since clinical measurements are done using electrodes, and the Complete Electrode Model \cite{Somersalo1992} would be more accurate. However, this is an initial feasibility study for comparing black-box and RGB machine learning approaches for stroke-EIT, and for this purpose a two-dimensional continuum model is sufficient. By black-box approach we mean a machine learning algorithm that learns directly from the raw data, without taking into account the specific inverse problems model. Our numerical experiments  suggest  that RGB inversion offers significantly better classification results than the black-box learning approach.

This paper is organized as follows. In Section \ref{sec:VHED_th}, we introduce the Virtual Hybrid Edge Detection method. Section \ref{sec:reconstr} gives the details of simulation of the EIT data and the VHED functions. Section \ref{section:Data} gives the details of the two neural networks, Fully Connected Neural Network and the Convolutional Neural Networks along with the training and testing data sets used for the classification. We report the numerical results of the classification of the two kinds of stroke using neural networks in Section \ref{sec:results}. Finally, Section \ref{sec:Conclusions} lists the concluding remarks.

\section{Virtual hybrid edge detection}\label{sec:VHED_th}

The main theoretical tool that has been used to study the inverse conductivity problem, is a special family of solutions to the conductivity equation \eqref{eq:cond}. These are called complex geometrical optics (CGO) solutions, or exponentially growing solutions, since they are characterized by a precise exponential asymptotic behavior at infinity. They have been first studied in inverse boundary value problems in the seminal paper \cite{sylvester1987}, but have appeared earlier in inverse scattering theory \cite{faddeev2016increasing, beals1984scattering}.

In the two dimensional case, for $L^\infty$ conductivities, Astala-P\"aiv\"arinta \cite{astala2006calderon} transformed the construction of the CGO solutions by reducing the conductivity equation to a Beltrami equation by setting $z = x_1 +i x_2$ and defining the Beltrami coefficient

\begin{equation*}
\mu(z) = \frac{1-\sigma(z)}{1+ \sigma(z)}
\end{equation*}

Since $c_1 \leq \sigma(z) \leq c_2$, we have $|\mu(z)| \leq 1 - \epsilon$ for some $\epsilon >0$. Further, assume $\sigma = 1$ outside of some $\Omega_0 \subset \Omega$, then $\text{supp}(\mu) \in \overline{\Omega_0}$.

Now, for $k \in \C$, and $ z \in  \C$, we define CGO solution as the unique solution of 
\begin{equation} \label{bel}
\overline{\partial}_{z} f_{\pm} (z,k) = \pm \mu (z) \overline{\partial_{z} f_{\pm}(z,k)}, \quad \quad
e^{-ikz}f_{\pm}(z,k) = 1+ \omega^{\pm} (z,k)
\end{equation}
where $ikz = ik(x_1+ix_2)$ and $\omega^{\pm}(z,k) = O(1/|z|)$ as $|z| \rightarrow \infty$. Here $z$ is considered as a spatial variable and $k$ as a spectral parameter.

A modified derivation of $\omega^{\pm}(z,k)$, which avoids the exponential growth and is computationally efficient was introduced in \cite{huhtanen2012numerical}.

A novel method of combing two imaging modalities of EIT and {\it{Virtual X-ray}} called Virtual Hybrid Edge Detection (VHED) was introduced in \cite{greenleaf2018propagation}. It was shown that, the leading term of a Born series derived from the boundary electrical measurements of EIT is a non-linear Radon transform of the conductivity, $\sigma$ and allows for a good reconstruction of the singularities of the conductivity. The method described in the paper was found to be very useful to detect nested inclusions within an inhomogeneous background conductivity. This is due to the fact that the well-posedness of Radon inversion results in robust method for detecting the leading singularities of the conductivity $\sigma$. A brief description of the method described in \cite{greenleaf2018propagation} is given below.

The Beltrami equation \eqref{bel} is treated as a scattering equation, with $\mu$ as a compactly supported scatterer and the incident field as $1$. The modified derivation of $\omega^\pm(z,k)$ as in \cite{huhtanen2012numerical}, then can be expanded as follows:
\begin{equation}
    \omega^\pm (z,k)= \sum_{n=1}^\infty \omega_n^{\pm}(z,k).
\end{equation}

 The first-order term $\omega_1 = \omega^+_1$ of the above Neumann series contains information allowing a stable edge and singularity detection of the coefficient $\mu$, and thus of the conductivity $\sigma$, the details of which are explained below.

Given the Neumann series (scattering) expansion of the CGO remainder term $\omega$, we consider the following transformations:
\begin{enumerate}[(i)]
    \item  First, one introduces polar coordinates in the complex frequency, $k$, writing $k=\tau e^{i\phi}$, with $\tau\in\R$ and $e^{i\phi}\in\sone$.
    
    \item Secondly, one takes a partial Fourier transform in $\tau$,  introducing a  nonphysical artificial (i.e.,  {\it virtual})  variable, $t$. This allows for good propagation of singularities from the interior of $\Omega$ to the boundary, allowing singularities of the conductivity in the interior to be robustly detected by voltage-current measurements at the boundary. 
\end{enumerate}

The Fourier transform of the $n-$th order scattering term in $\tau$ is denoted by 

\begin{equation}\label{VHED_official}
    \widehat{\omega}_{n}^{\pm}(z,t,e^{i\phi}):= \mathcal{F}_{\tau \rightarrow t}(\omega_{n}^{\pm} (z,\tau e^{i\phi})) = \int_{-\infty}^{\infty} e^{-it\tau} \omega_n^{\pm}(z,\tau e^{i\phi}) d \tau.
\end{equation}

where $t$ is a pseudo-time variable. The term $\widehat{\omega}^{\pm}(z,t,e^{i\phi})$ will be referred to as the {\it{VHED functions}} in this paper.

Consider now the following operators which apply a complex average to the first scattering term 
\begin{equation}\label{def:T1a}
    T_1^a: \mu(z_1) \mapsto \widehat \omega_1^{a}(t,e^{i\phi}) =\frac{1}{\pi} \int_{\partial \Omega}\widehat \omega_1^{+}(z,t,e^{i\phi})dz,
\end{equation}
and to the full scattering series
\begin{equation}\label{def:Tpm}
    T^\pm: \mu(z_1) \mapsto \widehat \omega^{a,\pm}(t,e^{i\phi}) =\frac{1}{\pi} \int_{\partial \Omega}\widehat \omega^\pm(z,t,e^{i\phi})dz.
\end{equation}

In \cite[\S 5B]{greenleaf2018propagation} it is shown that the following filtered-backprojection inversion formula holds:
\begin{equation}\label{eq:filtback}
    (T_1^a)^* (|D_t|^{-1})T_1^a = I \quad \text{on } L^2(\Omega),
\end{equation}
and
\begin{equation}\label{t0a expl}
T_1^a=-\big(\frac{i \pi }{\sqrt2}\big) e^{i\vp} \big(\frac{\partial}{\partial s}R\mu\big)(\frac12 t,e^{i\vp}),
\end{equation}
where $R$ is the standard Radon transform, 
$$(Rf)(s,\upsilon)=\int_{\mathbf x\cdot\upsilon=s}f(\mathbf x)\, d^1\mathbf x,\,\quad (s,\upsilon)\in\R\times \mathbb S^1.$$

Note that if we take $\Omega=\mathbb D$, where $\mathbb{D}$ is a unit circle,  
so that $\partial\Omega$ can be parametrized by $z=e^{i\theta}$, then (\ref{def:T1a}) becomes
\begin{equation}\label{omega circle}
\widehat \omega_1^a(t,e^{i\phi})= \int_0^{2\pi} \widehat \omega^+_1(e^{i\theta},t,e^{i\phi})\, ie^{i\theta}\,d\theta.\nonumber
\end{equation}

\begin{figure}
    \centering
\begin{picture}(320,250) 
\put(-65,100){\includegraphics[width=5.88cm]{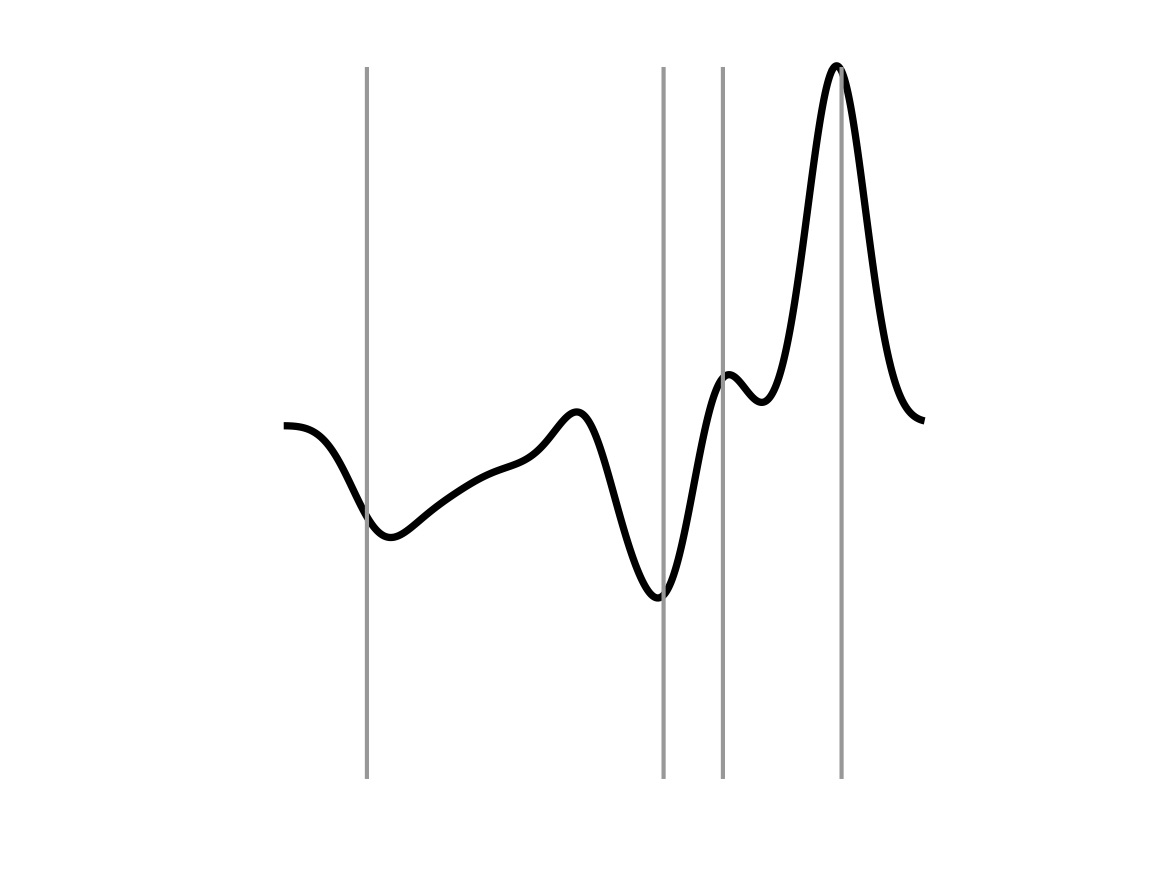}}
\put(65,100){\includegraphics[width=5.9cm]{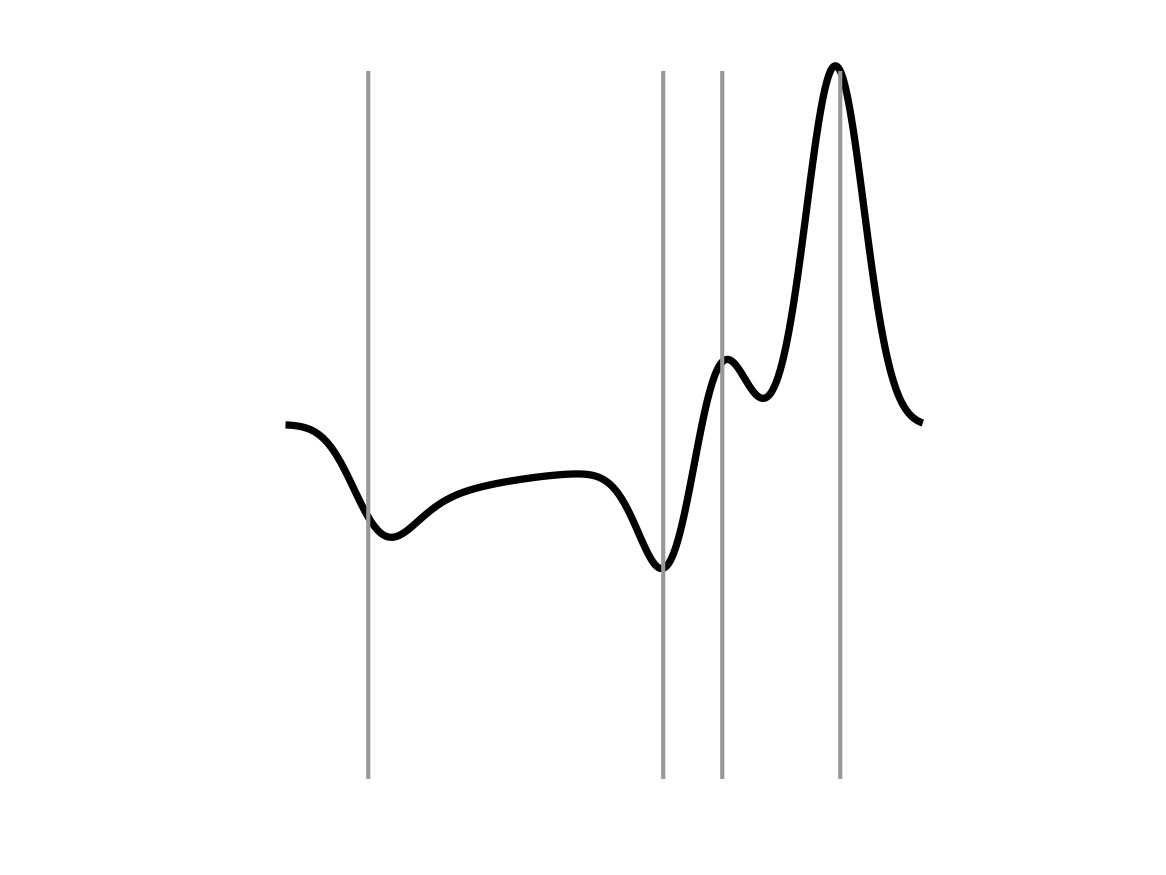}}
\put(202.5,100){\includegraphics[width=5.4cm,height=4.4cm]{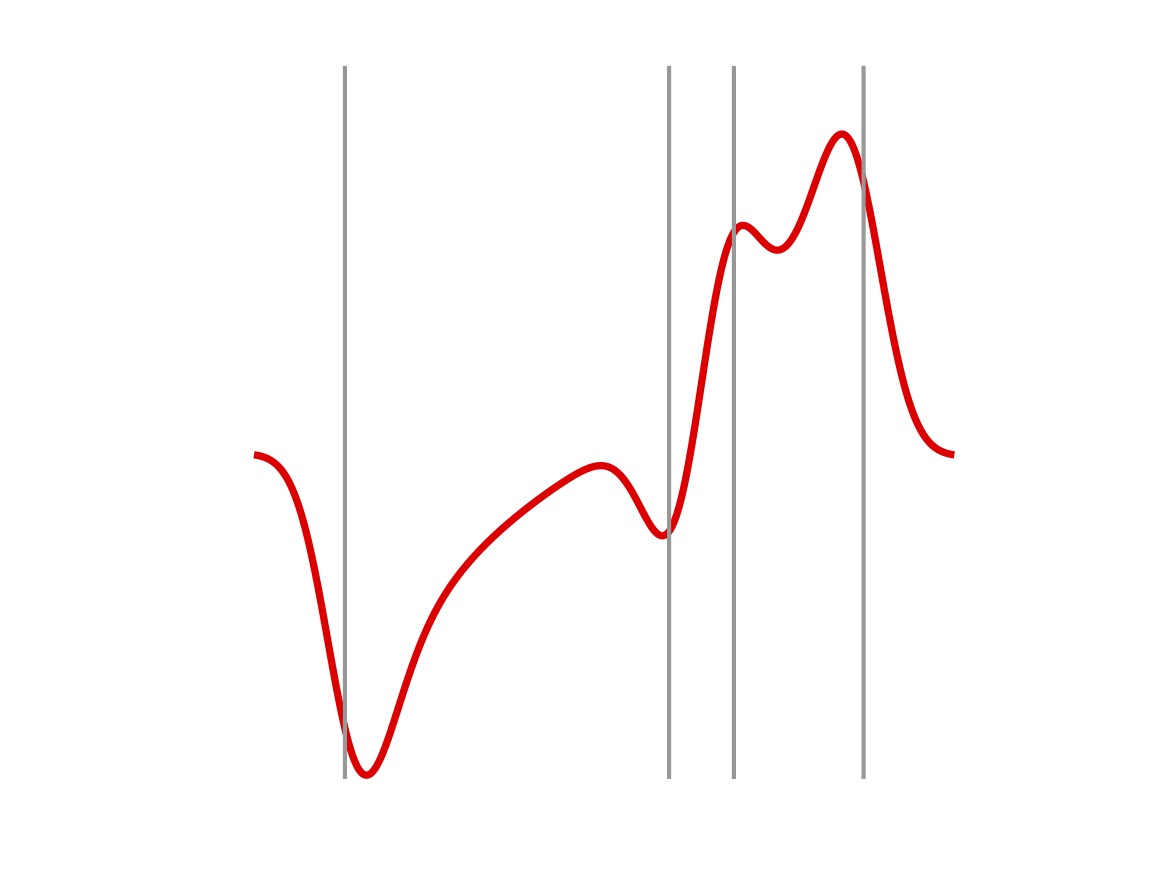}}
\put(-50,5){\includegraphics[width=4.9cm]{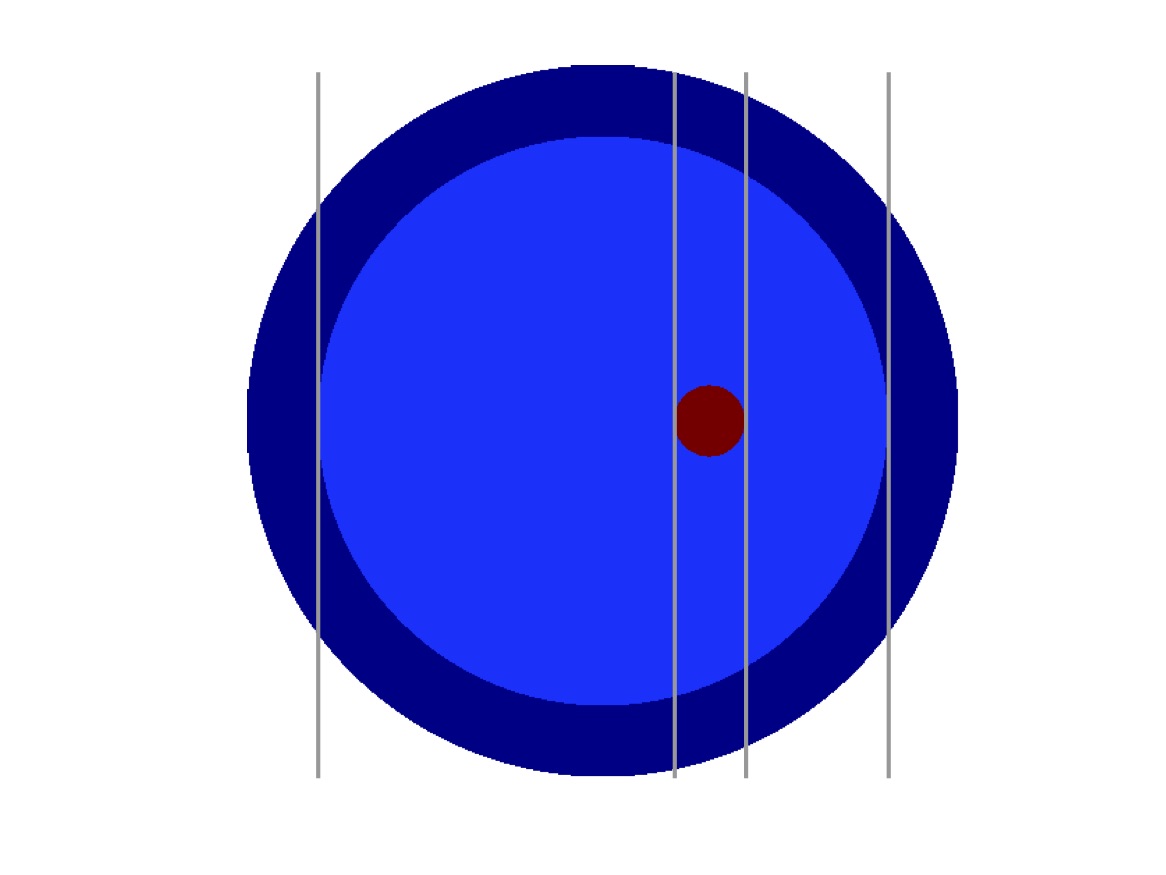}}
\put(80,5){\includegraphics[width=4.9cm]{cond_profile.jpg}}
\put(210,5){\includegraphics[width=4.9cm]{cond_profile.jpg}}
\put(-15,230){\Large{$\widehat{\omega}^+(1,2t,\!1)$}}
\put(90,230){\Large{$[\widehat{\omega}^+ \!\!-\!\widehat{\omega}^-](1,\!2t,\!1)$}}
\put(230,230){\Large \textcolor{red}{{$[T^{+}\!\!-\!\!T^{-}] \mu (2t,\!1)$}}}
\put(0,180){\vector(1,-1){13}}
\end{picture}    \caption{VHED functions for a simple conductivity having a small inclusion inside a larger inclusion. The grey lines indicate how the VHED functions carry information about specific parts of the wave front set of the conductivity. Actually the grey lines indicate the direction of the virtual X-rays in VHED. Left: just the plain VHED function $\widehat{\omega}^+(1,2t,\!1)$, featuring a nonlinear artifact, indicated by an arrow, in the center. Middle: subtracting the two kinds of VHED functions cancels the artifact in the center. However, the singularities at the leftmost and rightmost vertical grey lines have different amplitudes due to the choice $z=1$. Right: applying an averaging operator analogous to (\ref{def:T1a}) leads to equal-size (but opposite) singularities at the leftmost and rightmost vertical grey lines. This is because the $T^\pm$ operators integrate over $z\in\partial\Omega$.}
    \label{fig:VHED_illustration_1}
\end{figure}

Formula \eqref{t0a expl} provides an explicit and exact way to recover $\mu$ and its singularities, and thus for $\sigma$, from $\omega_1|_{\partial \Omega}$. We call the method Virtual Hybrid Edge Detection (VHED) since we are combining two imaging modalities, EIT and X-ray tomography, though the latter one is \textit{virtual}, since there are no actual X-rays involved. Figure \ref{fig:VHED_illustration_1} shows an example of VHED functions, before and after averaging, for a discontinuous conductivity on the unit disk. 

Unfortunately, the DN map only allows us to recover the boundary trace of the full Neumann series $\omega^\pm|_{\partial \Omega}$, as we will see later in Section~\ref{sec:reconstr}. In practice, though, the other terms do not contribute to the singularity detection and the exact inversion \eqref{eq:filtback} applied to $\omega^\pm|_{\partial \Omega}$ yields a good approximate reconstruction of $\mu$.

We recall that in order to precisely recover the singularity of $\sigma$, it is required to compute the Fourier transform of $\omega^\pm(z,k)$ in $\tau$.

With realistic and noisy data, we can compute $\omega^{\pm}(z,k)$ only in the disc $|k| \leq k_{\text{max}}$ with a measurement apparatus and noise independent radius $k_{\text{max}}>0$; see \cite{knudsen2009regularized} and the plots in Figure \ref{fig:VHED_noisedemo}. With smaller noise we can take a larger $k_{\text{max}}$; however, large noise forces $k_{\text{max}}$ to be small. This makes it more difficult to locate the singularities precisely.

\section{Simulating EIT data and VHED functions}\label{sec:reconstr}

In this paper we will compare machine learning methods based on two kinds of inputs:  (i) EIT data in the form of the continuum-model DN matrix, and (ii) the VHED functions. For the  computational experiments we need to simulate both of them for large collections of  test conductivities. We describe the numerical methods used for computing the DN matrix and the VHED functions in the following sections.

\subsection{Simulation of DN matrices}\label{sec:DN}

The computations of the Dirichelet to Neumann matrix (DN) and Neumann to Dirichlet matrix (ND) used for the machine learning algorithms in this paper follow along the same line as described in \cite{mueller2012linear}. 
Consider the trigonometric bases functions given by
\[  \phi_n(\theta) = \left \{
\begin{array}{c} 
      \pi^{-\frac{1}{2}} \text{cos} \bigg(\frac{(n+1) \theta}{2}\bigg), \quad n \quad \text{odd}  \\
      \pi^{-\frac{1}{2}} \text{sin} \bigg(\frac{n \theta}{2}\bigg),   \quad n \quad \text{even}\\ 
\end{array} \right. \]
where $\theta \in [0,2\pi]$ and $ n = \{1,2,3,...\}$. The constant basis function is given by $\phi_0(\theta)=(2\pi)^{-\frac{1}{2}}$.

The matrix approximation  ${\bf{L}}_{\sigma}$ of the DN map $\Lambda_\sigma$ is computed by solving the Dirichlet boundary problem
\begin{equation}\label{eq:D_p}
\nabla \cdot \sigma \nabla u_n = 0 \quad \text{ in } \Omega, \qquad  u_n|_{\partial \Omega}  = \phi_n
\end{equation}
and calculating the inner products 
	\begin{equation}\label{DNmap}
{\bf L}_{\sigma}:=[({\bf L}_{\sigma})_{m,n}] = \langle u_n|_{\partial \Omega}, \phi_m \rangle, 
	\end{equation} 
for row index $0 \leq m \leq 32$  and column index $0 \leq n \leq 32$.

However, in practice, we form the DN matrix ${\bf{L}}_{\sigma}$ by first computing the ND matrix ${\bf{R}}_{\sigma}$. This is because, calculating the ND matrix ${\bf{R}}_{\sigma}$ does not involve the somewhat unstable step of numerical differentiation. To compute the ND matrix ${\bf{R}}_{\sigma}$, we consider the Neumann problem 
\begin{equation}\label{eq:cond2}
\nabla \cdot \sigma \nabla v_n = 0 \quad \text{ in } \Omega, \qquad  \sigma \frac{\partial v_n}{\partial \nu}\bigg|_{\partial \Omega} = \phi_n 
\end{equation}
where the solution $v_n$ is made unique by the requirement $\int_{\partial \Omega} v_n dS = 0$. 

We use the Finite Element Method with a mesh of $131072$ triangle elements to solve equation~(\ref{eq:cond2}) numerically and recover the boundary voltages, $v_n|_{\partial \Omega}$. We then compute the matrix approximation of ND map ${\bf{R}}_{\sigma}$ using
	\begin{equation}\label{NDmap2}
{\bf R}_{\sigma}:=[({\bf R}_{\sigma})_{m,n}] = \langle v_n|_{\partial \Omega}, \phi_m \rangle, 
	\end{equation}
where $1 \leq m \leq 32$ is the row index and $1 \leq n \leq 32$ is the column index. We avoid the the constant basis function $\phi_0 = \frac{1}{\sqrt{2 \pi}}$ for the ND matrix ${\bf{R}}_{\sigma}$, since Neumann data must satisfy $\int_{\partial \Omega} \sigma \frac{\partial v_n}{\partial \nu} dS = 0$. 

Now in the linear subspace, $\text{span}\{\phi_1,\phi_2,\ldots \phi_{32}\}$ we have ${\bf{L}}_{\sigma} = {\bf{R}}_{\sigma}^{-1}$.
We extend the DN matrix onto the full space, 
$\text{span}\{\phi_0, \phi_1, \ldots \phi_{32} \}$
as follows:
\begin{equation*}
    {\bf{L}}_{\sigma} = \left [ \begin{array}{c c}
    0 & 0 \\
    0 & {\bf{R}}_{\sigma}^{-1} \end{array}\right]
\end{equation*}

In this paper we use noisy DN matrix for the computation of the VHED functions and for the classification using neural networks. We compute the noisy DN matrix as follows:

Add zero-mean random Gaussian noise to the DN matrix ${\bf {L}}_{\sigma}$ to obtain a noisy matrix
${\bf {L}}_{\sigma}^{\delta}$, with relative noise given by
\begin{equation}\label{Noisy_DN}
\delta  = \frac{ \| {\bf {L}}_{\sigma}^{\delta} -  {\bf{L}}_{\sigma} \|_{2}} {\|{\bf{L}}_{\sigma}\|_{2}},
\end{equation}
where $\|\cdot\|_2$ denotes the spectral norm of the matrix. 
We add a relative noise of $\delta = 10^{-3}$ and $\delta = 10^{-2}$ to the DN matrix, ${\bf{L}}_{\sigma}$ to get ${\bf {L}}_{\sigma}^{\delta}$. We use both ${\bf{L}}_{\sigma}$ and ${\bf {L}}_{\sigma}^{\delta}$ for the computation of VHED functions and as an input to the Neural Network.
 
\subsection{Simulation of VHED profiles}\label{sec:VHED}

We wish to evaluate numerically VHED functions  $\widehat \omega^\pm (z,t,e^{i\phi})$ for a selection of boundary points $z \in \partial \Omega$, pseudo-times $t \in \R$ and virtual X-ray directions $\phi \in [0,2\pi]$. We recall from \cite[Section 10A]{greenleaf2018propagation} the steps to construct the VHED profiles from the DN map, $\Lambda_\sigma$.

\begin{enumerate}[(i).] 

\item In order to construct the CGO solutions $\omega^{\pm}(z,k)$  from the DN map $\Lambda_{\sigma}$, we use a $\mu$ - Hilbert transformation $\Hm_{\mu}$ satisfying
\begin{equation*}
    \partial_{T} \Hm_{\mu} f = \Lambda _{\sigma} f
\end{equation*}
 and 
 \begin{equation*}
 \int_{\partial \Omega} \Hm_{\mu} f dS = 0
 \end{equation*} where the $\partial_{T}$ represents the tangential derivative and $f \in H^{1/2}(\partial \Omega)$.

\item Defining the averaging operator as follows: 
$$\mathcal{L} \varphi = |\partial \Omega|^{-1} \int_{\partial \Omega} \varphi\, ds $$ and
the real-linear operator
$$ \mathcal{P}_{\pm \mu} f = \frac{1}{2} (I +i \Hm_{\pm \mu})f +\frac{1}{2} \mathcal{L} f $$ where $f \in \mathbb{C}$.

\item This allows us to define the projection operator $\Pc_{\pm \mu}^{k}$ as
\begin{equation}\label{proj_op}
\Pc_{\pm \mu}^{k} := e^{-ikz} \mathcal{P}_{\pm \mu} e^{ikz}
\end{equation} 

\item We then find the CGO solution $f_{\pm \mu}(z,k) = e^{ikz}(1+\omega^{\pm}(z,k)) $, and thus $\omega^{\pm}(z,k)$, for $k \in \C$, by solving the boundary integral equation
\begin{equation}\label{BIE}
f_{\pm \mu}(z,k) +e^{ikz} = (\Pc_{\pm \mu}^{k} + \Pc_0^k)f_{\pm \mu}(z,k), \qquad z \in \partial \Omega,
\end{equation}
where $\Pc_{\pm \mu}^{k}$ and $\Pc_0^k$ are projection operators defined by \eqref{proj_op}.
\end{enumerate}

\noindent
If we had infinite-precision EIT data available, the solution of (\ref{BIE}) would be perfect, and VHED functions $\widehat{\omega}^{\pm}(z,t,e^{i\phi})$ could be computed with the one-dimensional Fourier transform as defined in (\ref{VHED_official}). 

\begin{figure}
    \centering
\begin{picture}(320,150)
\put(10,0){\includegraphics[width=9cm]{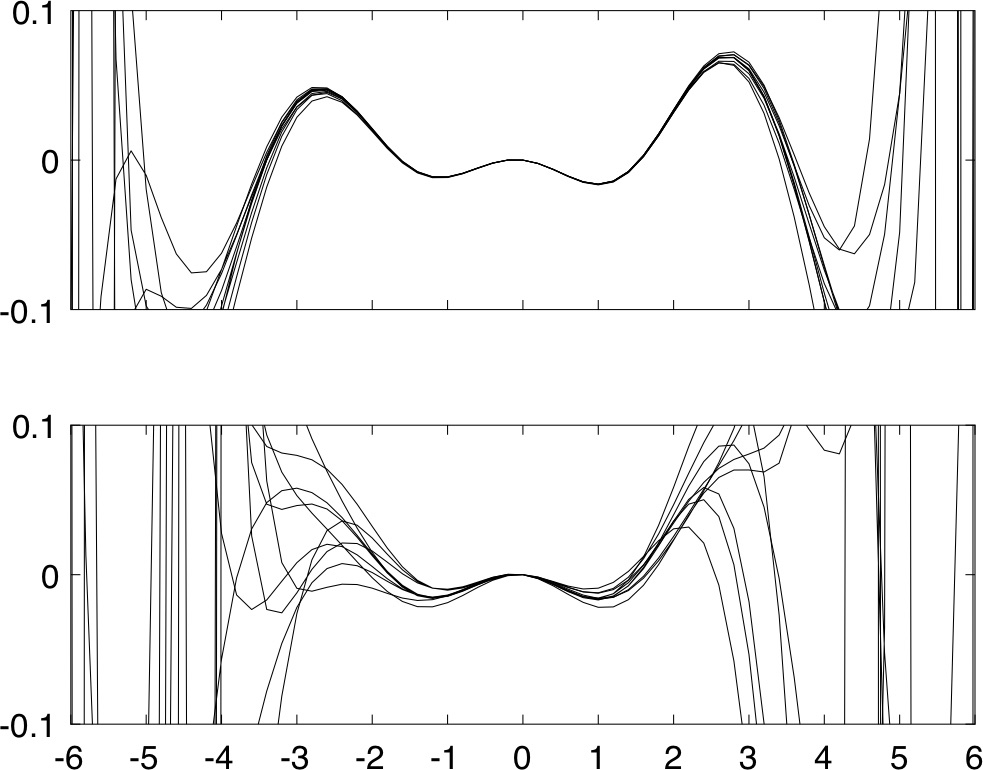}}
\end{picture}    
\caption{Illustration of the effect of noise on the CGO solutions computed via the boundary integral equation. Top plot: $\mbox{Re}(\omega^+(1,\tau e^{i\phi}))$ as a function of real-valued $\tau$ and with fixed $\phi=0$. We computed $\mbox{Re}(\omega^+(1,\tau e^{i\phi}))$ with ten different random realizations of noise added to the ND matrix. Relative noise amplitude is here 0.00005. The magnitude of error in the computations becomes larger as $|\tau|$ grows. Bottom: same as top, but higher noise amplitude 0.001. Note that stronger noise makes the stable interval around $\tau=0$ more narrow.}
    \label{fig:VHED_noisedemo}
\end{figure}

In practice, though, measurement noise in the DN matrix causes inaccuracy in the solution of the boundary integral equation (\ref{BIE}). Moreover, the errors in the numerical point values of traces  $\omega^{\pm}(z,\tau e^{i\phi})$ for $z\in\partial\Omega$ become exponentially larger as $|\tau|$ grows; see Figure \ref{fig:VHED_noisedemo} for an illustration. 

So for practical computations we use the windowed Fourier transform:
\begin{equation}\label{win_FT}
    \widehat{\omega}^{\pm}_R (z,t,e^{i\phi}) = \int_{-R}^{R} W_{a_R}(\tau)e^{-it \tau}\omega^{\pm}(z,\tau e^{i\phi}) d \tau.
\end{equation}
Compared to (\ref{VHED_official}), formula (\ref{win_FT}) introduces two new things: the cutoff frequency $R>0$ and the window function  $W_{a_R}:\R\rightarrow\R$ given by
\begin{equation}\label{windowfunction}
W_{a_R}(\tau) = \exp(-a_R\tau^2).
\end{equation}
In (\ref{windowfunction}) we choose  $a_R>0$ so large that $W_{a_R}(\tau)$ is very small outside the interval $\tau\in[-R,R]$. 

From the theory of the Fourier transform we know that windowing on the frequency domain corresponds to convolution in the pseudo-time domain:
$$
\widehat{\omega}^{\pm}_R (z,t,e^{i\phi}) = \widehat{W}_{a_R}(t)\ast \widehat{\omega}^{\pm} (z,t,e^{i\phi}).
$$
Therefore, the measurement noise forces us to work with blurred versions of the VHED functions, and the blurriness is more severe for stronger noise. See Figure \ref{fig:two_lowpass_filters_VHED}.

However, the windowing also gives us something valuable: noise-robustness. To see this, consider the exact windowed VHED function $\widehat{\omega}^{+}_R (1,t,1)$ and an approximation $\widehat{\omega}^{+,\delta}_R (1,t,1)$ of the same function computed from noisy data via the boundary integral equation (\ref{BIE}). We can then estimate the error between them by 

\begin{align}
    & \nonumber
    | \widehat{\omega}^{+}_R (1,t,1) - \widehat{\omega}^{+,\delta}_R (1,t,1) | \\
    &= \nonumber
    \left|\int_{-R}^{R} e^{-i \tau t} (\omega^+(1,\tau) - \omega^{+,\delta}(1,\tau) ) W_{a_R}(\tau) d \tau  \right| \\
    & \label{errorestimate}
    \leq \int_{-R}^{R}  |\omega^+(1,\tau) - \omega^{+,\delta}(1,\tau) | \, W_{a_R}(\tau) \, d \tau.   
\end{align}
Now the function $W_{a_R}(\tau) $ is small near the endpoints of the interval $\tau\in[-R,R]$ where the function $|\omega^+(1,\tau) - \omega^{+,\delta}(1,\tau) |$ is large (see Figure \ref{fig:VHED_noisedemo}), and vice versa. Therefore the right-hand side in (\ref{errorestimate}) is small.

In this paper we take the cutoff frequency to be $R =4$ and choose $a_{R}=0.35$. This choice restricts $W_{a_R}(\tau)$ to be very small outside the interval $\tau \in [-4,4]$. In this paper we restrict the {\textit{Virtual X-rays}} to one direction given by $\phi = 0$. Thus, the windowed VHED function used for the neural networks is given by $\widehat{\omega}^{\pm} (z,t,1)$, where we have dropped the subscript $W_{a_R}$. 

While it is conceivable that a human observer would learn to classify strokes based on the (unrealistic) data shown on the left in Figure \ref{fig:two_lowpass_filters_VHED}, we believe that machine learning is needed for the type of blurred (realistic) data shown on the right in Figure \ref{fig:two_lowpass_filters_VHED}.

\begin{figure}
    \centering
\begin{picture}(320,80)
\put(-45,-20){\includegraphics[width=5cm]{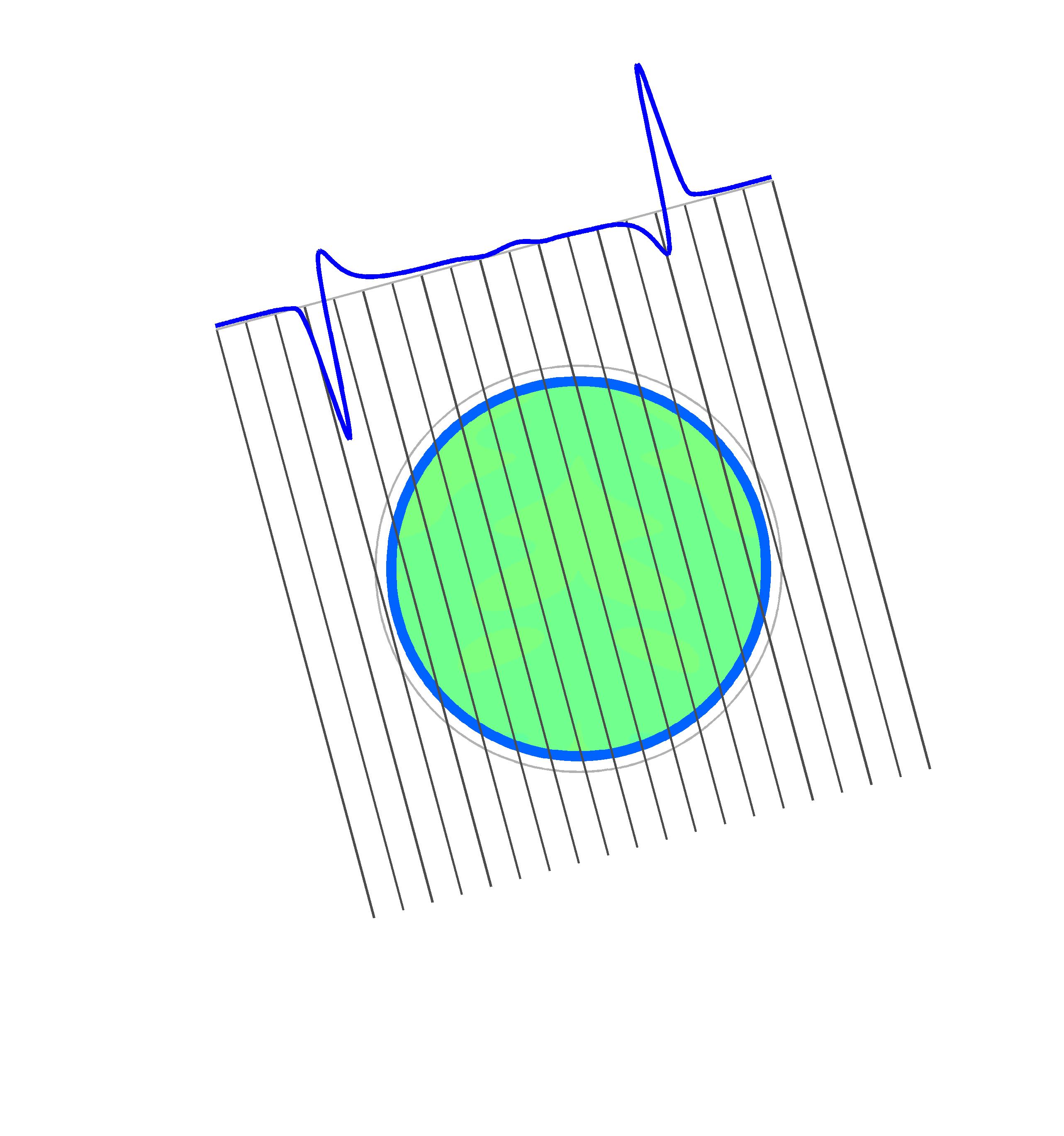}}
\put(80,-20){\includegraphics[width=5cm]{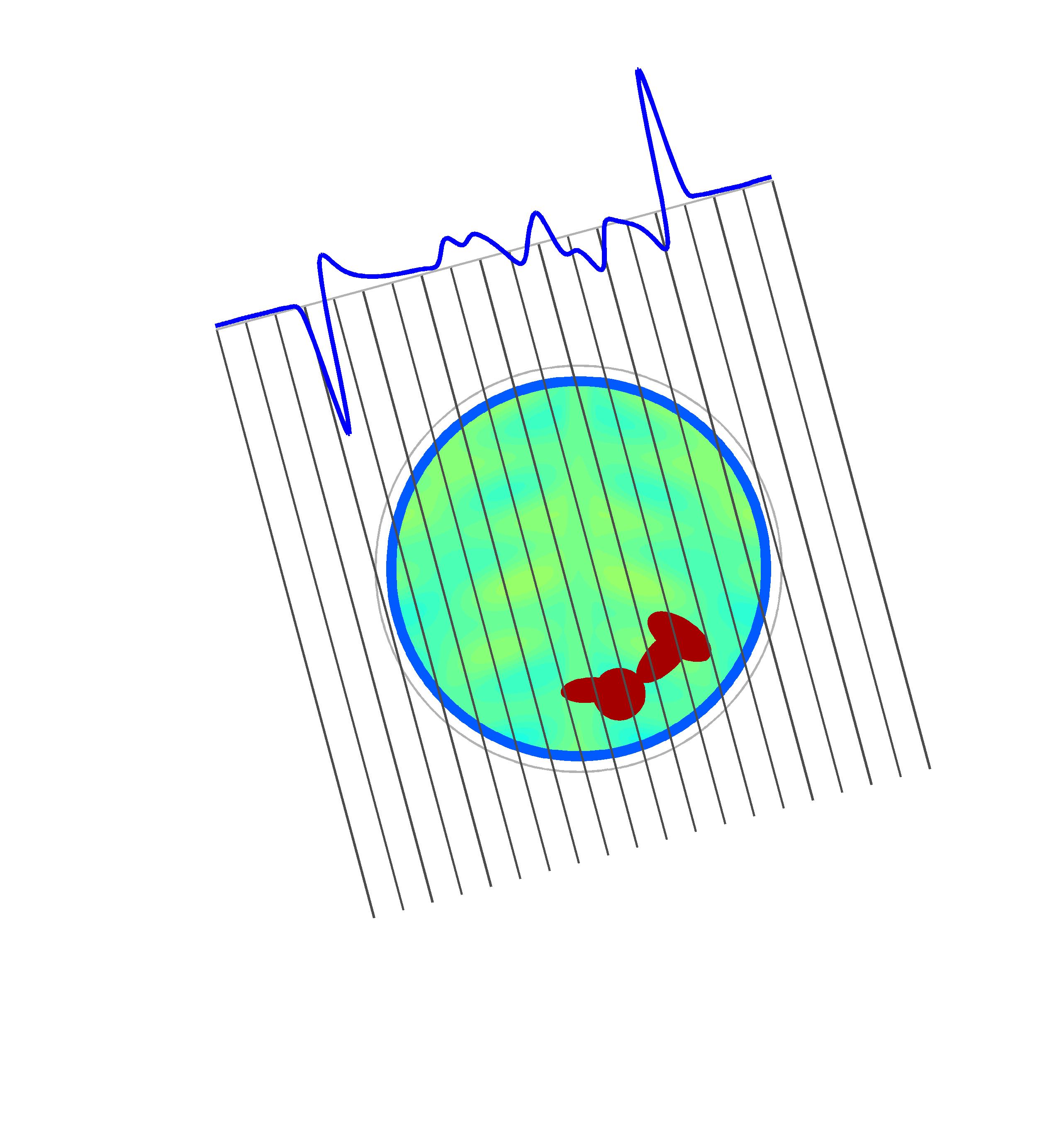}}
\put(205,-20){\includegraphics[width=5cm]{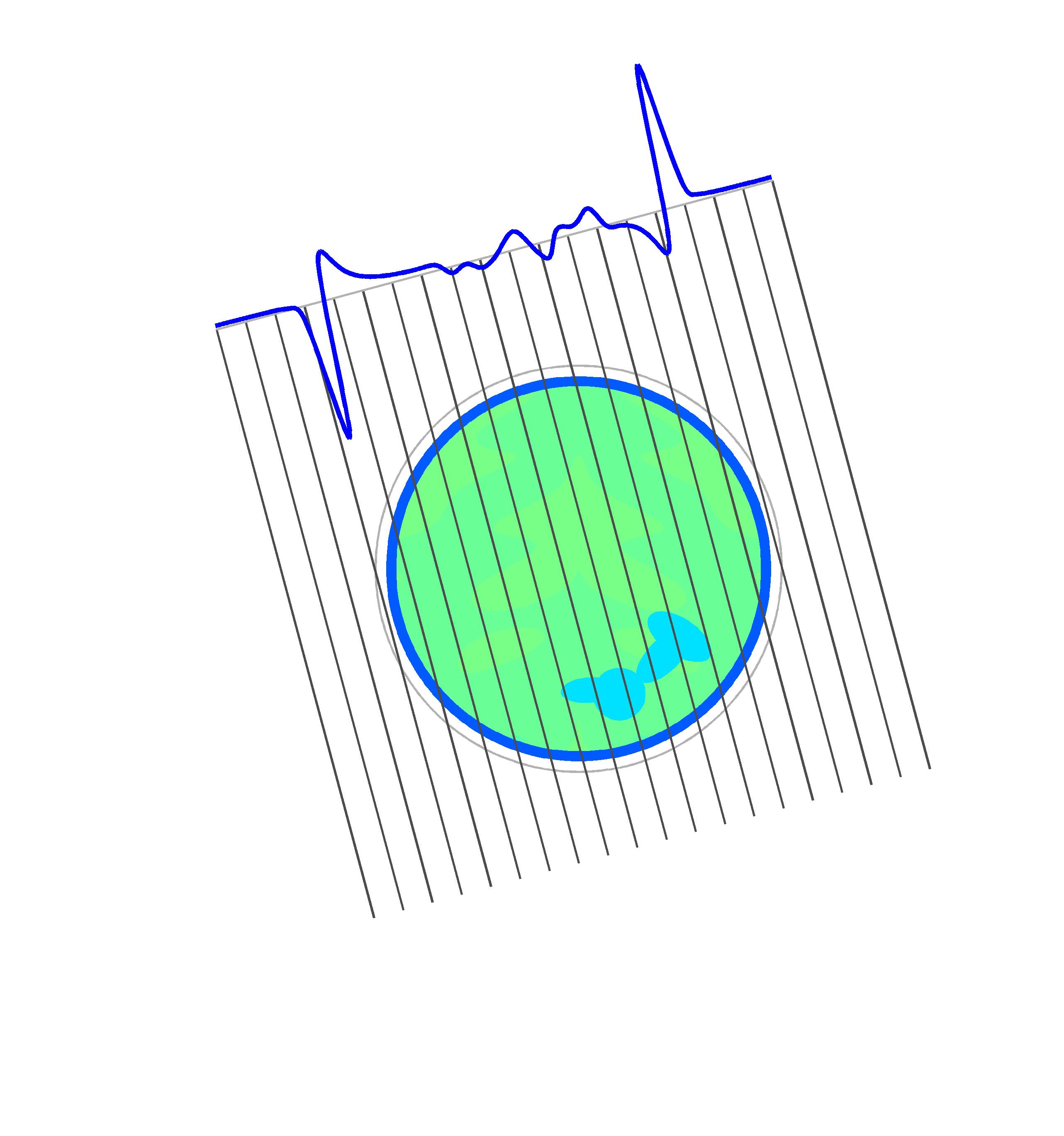}}
\end{picture}
    \caption{Left: a two-dimensional simulated model of a slice through a human head. Modelled are the resistive skull (blue annnulus) and slightly varying brain tissue (greenish). Virtual X-rays are shown as gray lines, and blue function plot is the integrated VHED function $[T^{+}\!\!-\!\!T^{-}] \mu (2t,\!1)$. The Fourier transform is windowed in the interval $[-60,60]$. Note how the discontinuities at the skull boundaries are reflected as sharp peaks in the VHED function. Middle: The same VHED function as on the left, but with a simulated hemorrhage in the brain. The red area (modelling the bleed) has higher conductivity than background. Right: The same VHED function as on the left, but with a simulated ischemia in the brain. The light blue area (modelling the part of the brain deprived of blood flow) has lower conductivity than background.}
    \label{fig:ThreeStrokeComparisonVHED}
\end{figure}

\begin{figure}
    \centering
\begin{picture}(320,100)
\put(-45,15){\includegraphics[width=7cm]{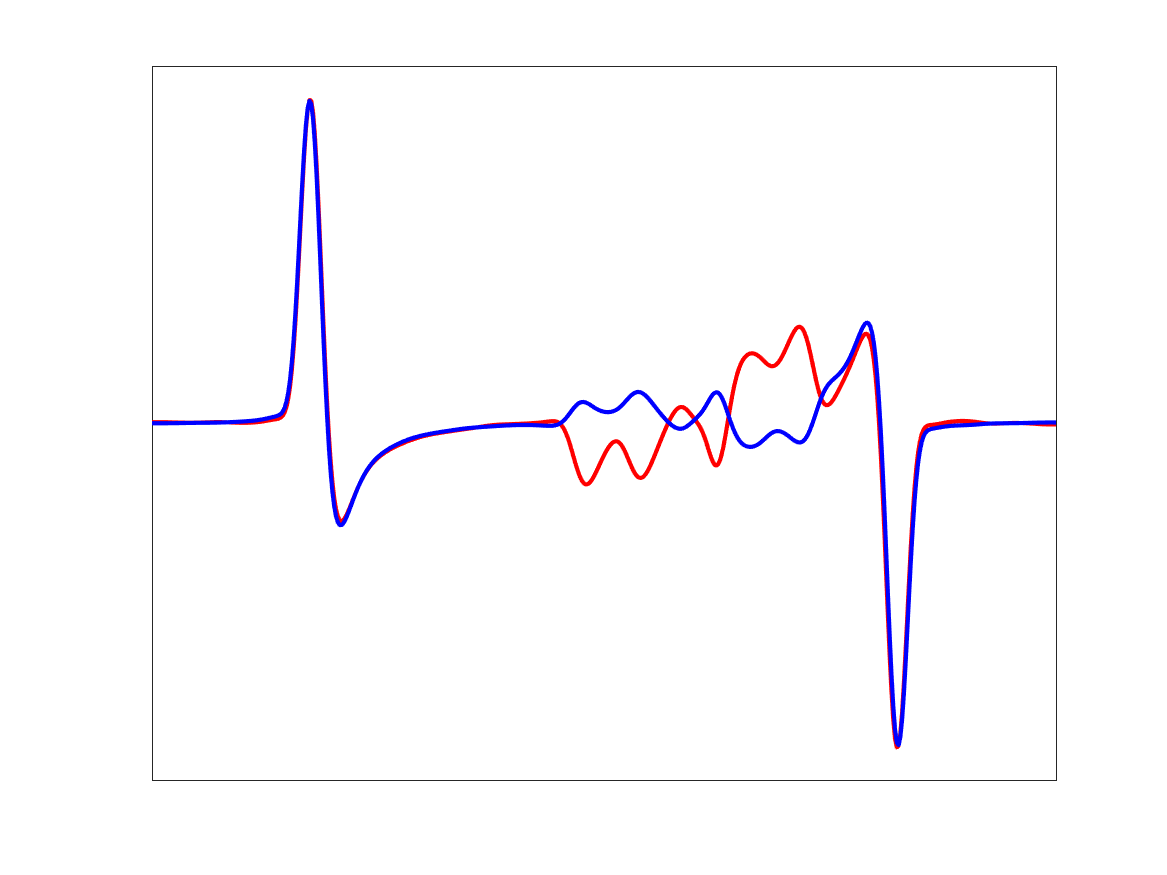}}
\put(-10,7){$\displaystyle \int_{-60}^{60} W_{a_{60}}(\tau)e^{-i t\tau }\omega^\pm(z,\tau,e^{i\varphi})\,d\tau$}
\put(145,15){\includegraphics[width=7cm]{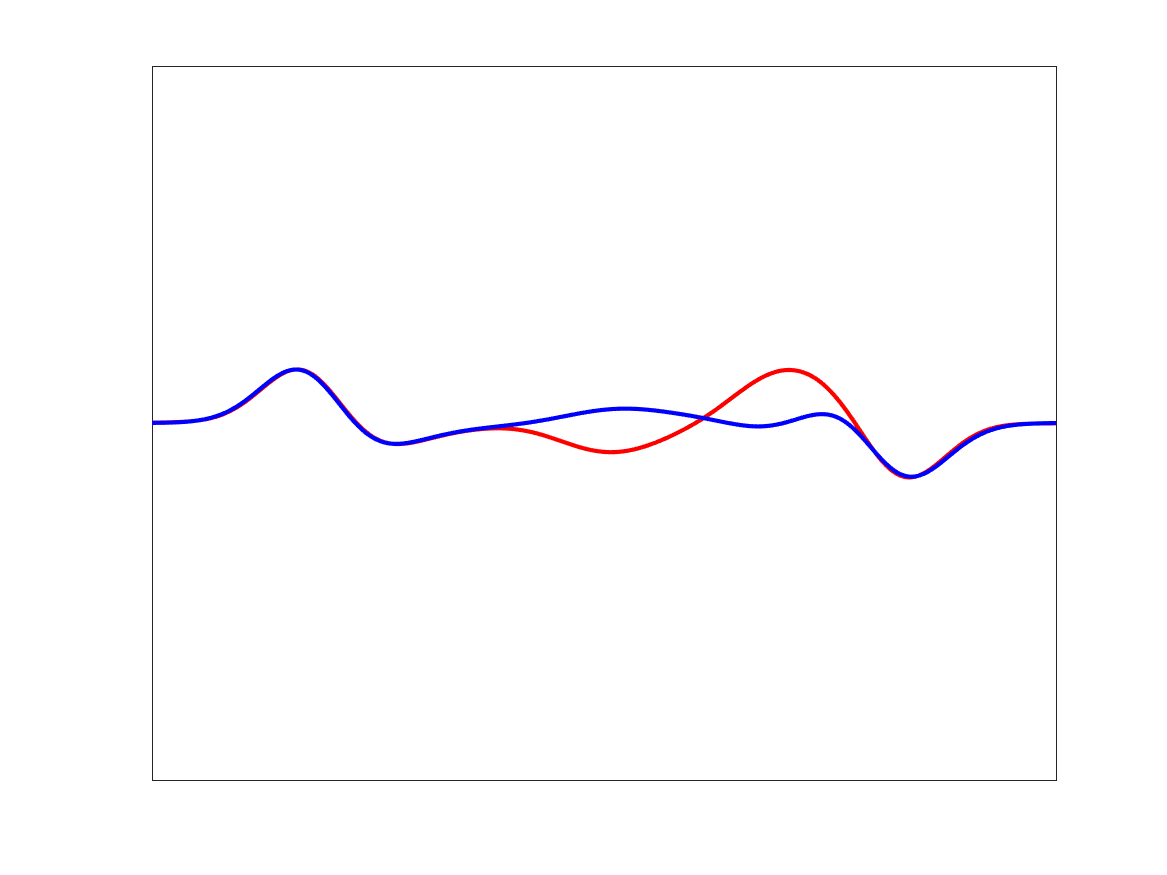}}
\put(180,7){$\displaystyle \int_{-4}^{4} W_{a_{4}}(\tau)e^{-i t\tau }\omega^\pm(z,\tau,e^{i\varphi})\,d\tau$}
\end{picture}    
\caption{Effect of low-pass filtering to the VHED functions shown in Figure \ref{fig:ThreeStrokeComparisonVHED}. Measurement noise prevents us from computing the CGO solutions outside a noise-dependent interval $\tau\in[-R,R]$.  Left: unrealistic VHED profiles with cutoff frequency $R=60$ for the two stroke examples in Figure \ref{fig:ThreeStrokeComparisonVHED} (red for hemorrhage and blue for ischemic). Right: same as left, but cutoff frequency $R=4$. Here $a_{4} = 0.35$ and $a_{60} = 0.0019$.}
    \label{fig:two_lowpass_filters_VHED}
\end{figure}

\section{Neural Networks}\label{section:Data}

Neural networks are data adaptable computational models commonly used in machine learning. They are often capable of capturing nonlinear relations between input-output pairs called the {\it training pair} and generalizing them to new inputs they have not encountered during training. Despite their great success in classification and regression tasks, identifying possible causal relationships between input and output in neural network models is an active research area and not completely understood as of now. Therefore, neural networks are still considered a \textit{blackbox} approach \cite{doshi2017towards}. 

Neural networks can be very sensitive to the input data. A slight noise in the input data sometimes makes a lot of difference in the prediction of the output. There are examples of this in the image processing and computer vision applications of neural networks \cite{lunz2018adversarial}.

The solution techniques for ill-posed inverse problems on the other hand have  well-established causal relationship between the input and output. Also, they are noise-robust because of regularization. Thus, they offer the two advantages with which the neural network approach often has trouble with. However, especially in EIT, regularization typically leads to blurring in reconstructed images. Therefore, combining methods of machine learning and traditional inversion mathematics might be a good combination. 

The goal of this study is to classify the two different kind of strokes, ischemic and hemorrhagic, using neural networks on simulated brain. To achieve this end, we make use of two different types of input data: The EIT data DN matrix, ${\bf{L}}_{\sigma}$, and the VHED function
\begin{equation}\label{VHED_def}
\widehat{\omega}(z,t,1) = \widehat{\omega}^{+}(z,t,1) - \widehat{\omega}^{-}(z,t,1).
\end{equation}
We use two neural networks, Fully Connected Neural Networks and Convolutional Neural Networks with two different inputs mentioned above to study the feasibility of the classification. In this study we show numerical results to the end that the neural networks are better capable of distinguishing the  type of stroke with VHED functions $\widehat\omega(z,t,1)$ as input rather than the EIT data, DN matrix ${\bf{L}}_{\sigma}$ as an input to the neural network.

\subsection{Training and Testing datasets}\label{section:TT_data}

\begin{figure}
 \centering
 \begin{picture}(540,150) 
 \put(-15,0){\includegraphics[width=7cm]{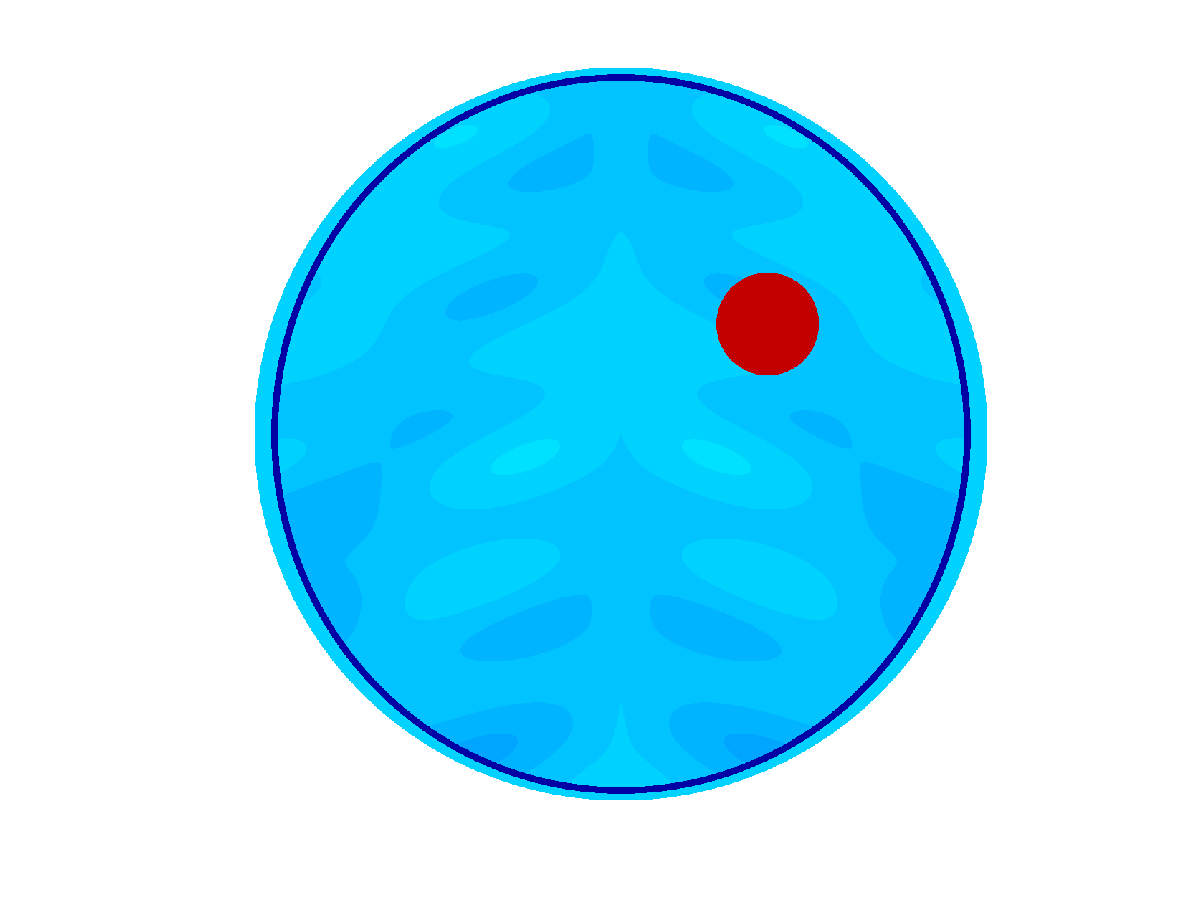}}
 \put(155,0){\includegraphics[width=7cm]{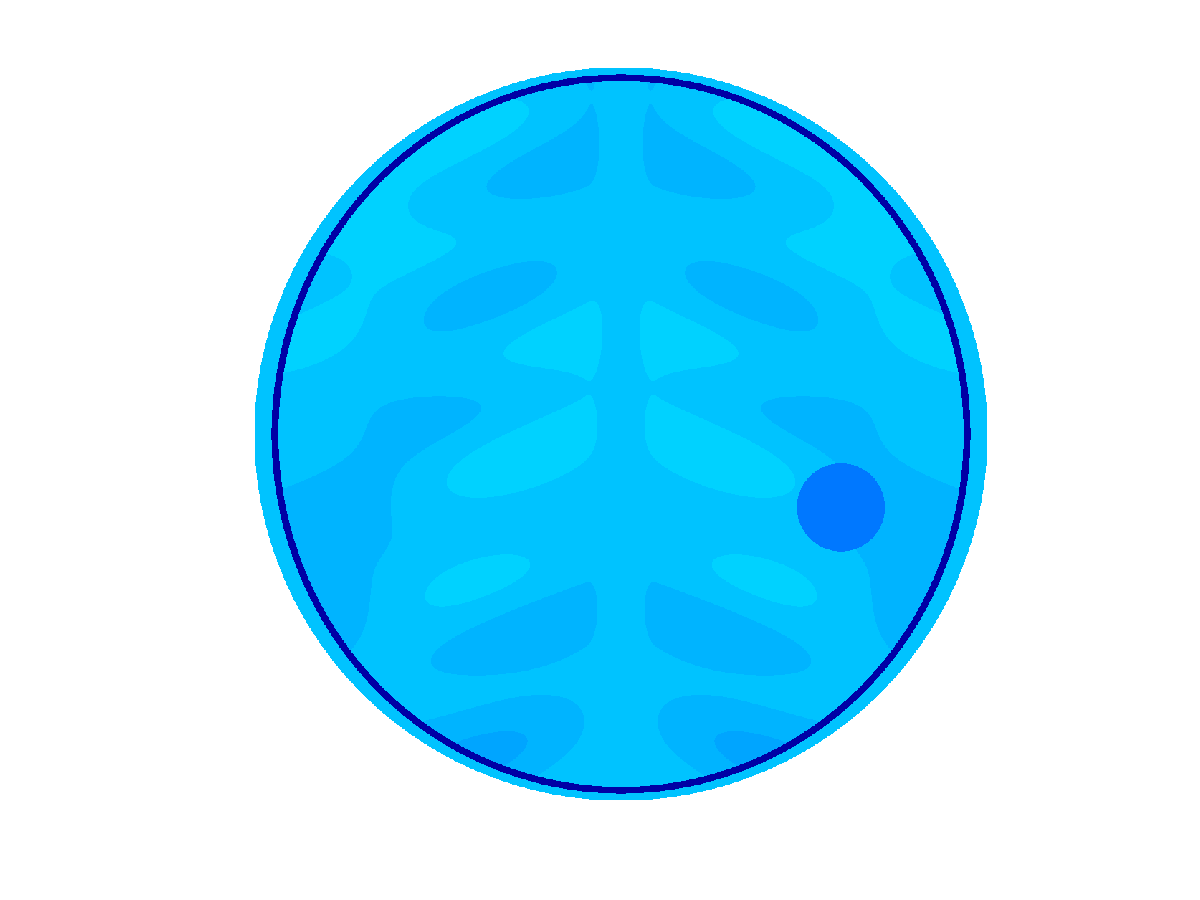}}
 \end{picture}    
 \caption{ A representation of the training data set. Left figure represents a hemorrhagic stroke. Right figure represents the ischemic stroke. The $2$-Dimensional cross section of the brain is assumed to be a unit disk. There are three distinct regions in the brain apart from the inclusion of the stroke. The outer layer representing scalp with a constant conductivity $1$. A middle layer representing skull with homogeneous conductivity that is elliptical in shape and the interior of the brain has inhomogeneous distribution of the conductivity. The  width of the skull, the conductivities of the various layers , and the area and location of the inclusion are sampled randomly from a uniform distribution in the ranges listed in the Table \ref{table:sim_data}. The stroke is a circular inclusion for the training data and is present on the right side of the disk.}
 \label{fig:training_set}
\end{figure}

The guiding example behind our simplified computational measurement model is a  band of electrodes attached linearly around the head of a patient.
The training and testing data set for the neural networks for this study are made of synthetic data, which is constructed as follows. 

We consider a simplified model of a two-dimensional cross-section of the human head. We use the so called continuum model on the one-dimensional boundary curve of the cross section. This involves two severe simplifications. First, we are ignoring the finite number of electrodes used in practice and instead study continuously applied voltage distributions. In the past EIT research, the step from continuum model to more accurate electrode models has been successfully taken, and we expect a future work to bring similar success to the present topic. 

Second, we approximate a three-dimensional structure with a two-dimensional model. This inevitably introduces some modelling error. However, in lung EIT this has been found to be acceptable \cite{Mueller2018}.

We consider a unit disc to model the $2$- dimensional cross section of the head. We construct an outer layer with constant conductivity of $1$ to represent the scalp. A middle layer that is elliptical with homogeneous conductivity represents the skull and an inner layer with inhomogeneous conductivity representing the grey and white matter of the brain.

In each of the unit discs, there is an inclusion representing the stroke on the right side of the disc. Why only on the right? Almost all strokes are confined to one half of the brain only, and the affected half is easily identifiable from the symptoms. For example, if the patient cannot lift their left hand up, then the stroke is in the right hemisphere. Therefore there is no loss of generality in placing all the simulated strokes on the right as we can always reflect the measurement data if needed. The inclusion representing the ischemic stroke has lower conductivity and inclusion representing the hemorrhagic stroke has higher conductivity than the background.

In each of these parameters such as width and conductivity of the skull, conductivity of the interior of the brain, area and location and conductivity of the inclusion are randomly sampled from a uniform distribution in the ranges given in Table \ref{table:sim_data}.

For the training data set, the inclusion is circular in shape. See Figure \ref{fig:training_set} for an example of the training data set. We consider four different options for the testing data set: (a) Circular inclusion (b) Elliptic Inclusion (c) An inclusion that is irregular in shape and (d) Multiple inclusions.

\begin{figure}
	\centering
	\begin{picture}(540,125) 
	\put(-0,105){\includegraphics[width=6.0cm]{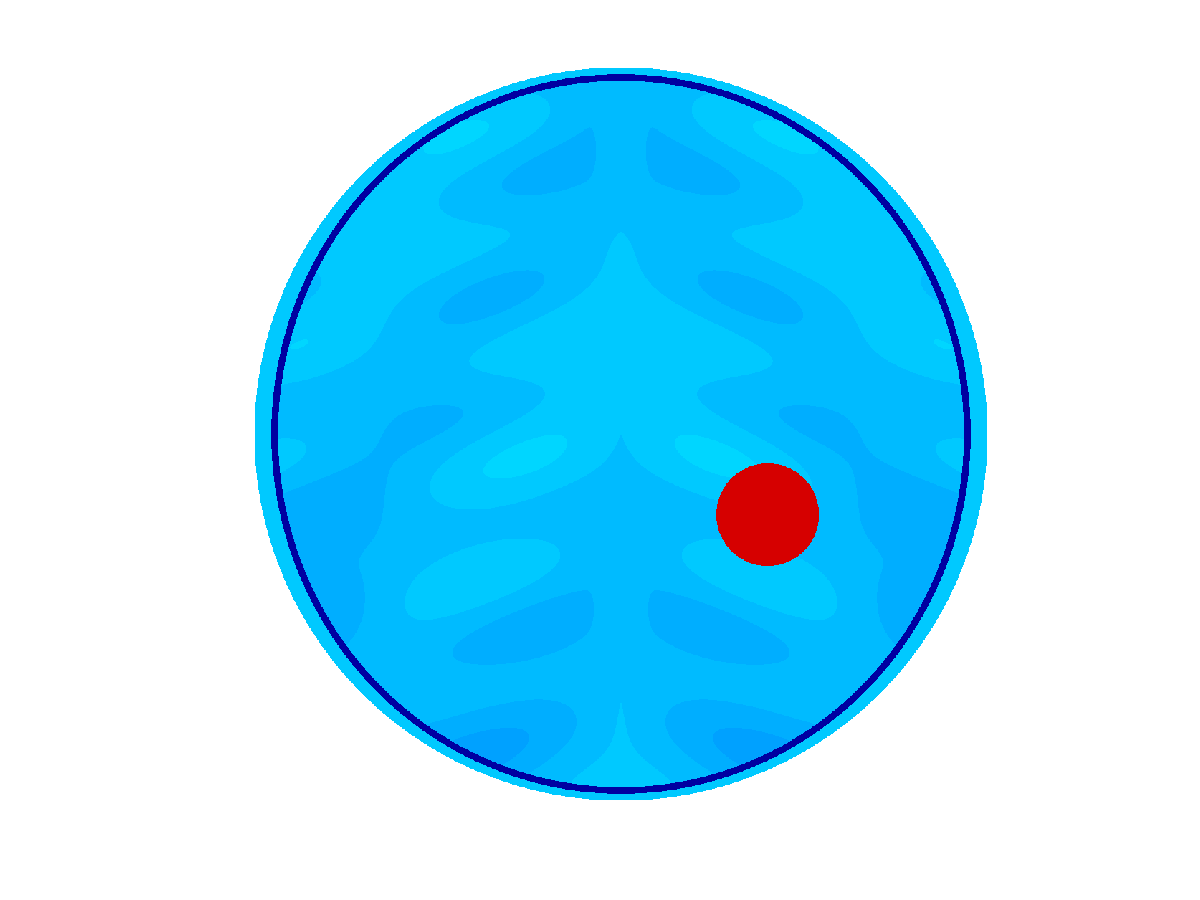}}
	\put(150,105) {\includegraphics[width=6.0cm]{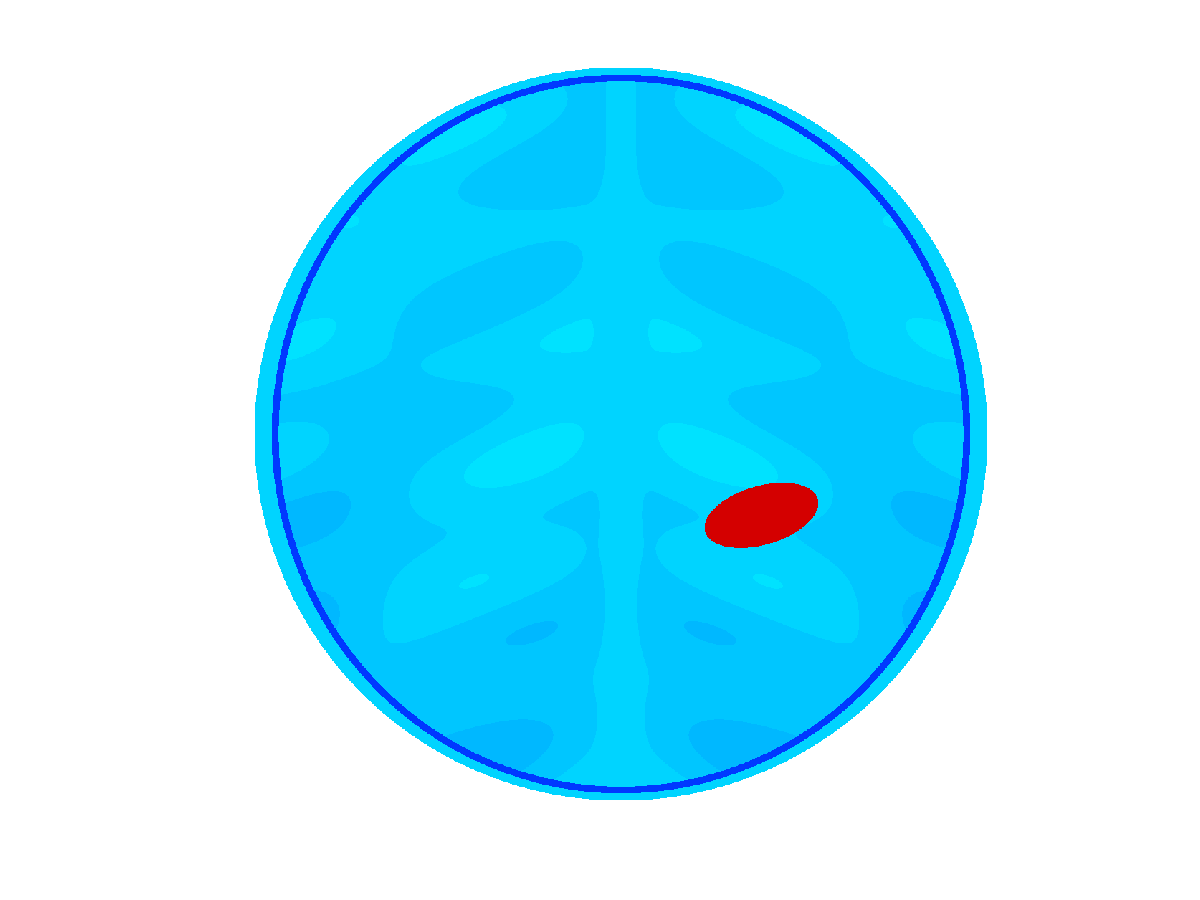}}
	\put(-0,-10){\includegraphics[width=6.0cm]{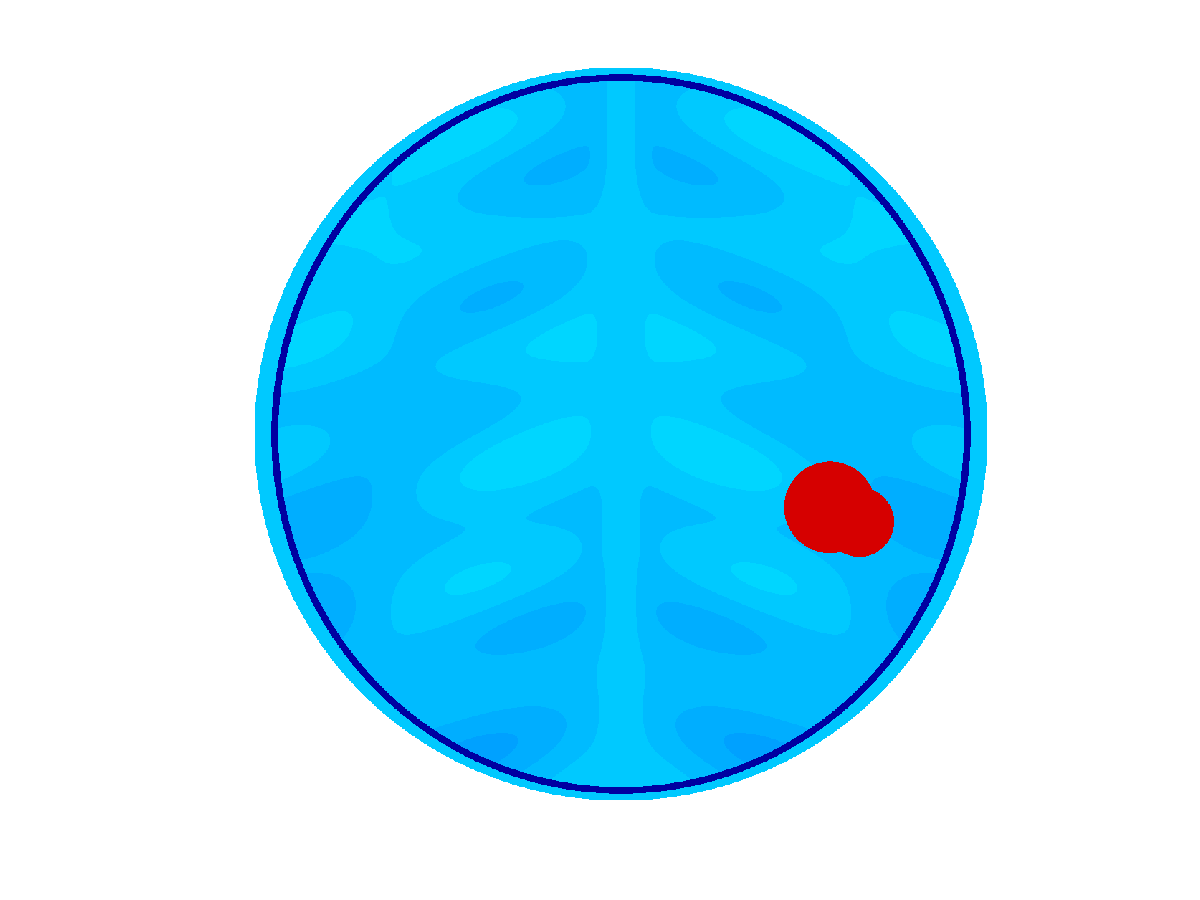}}
	\put(150,-10){\includegraphics[width=6.0cm]{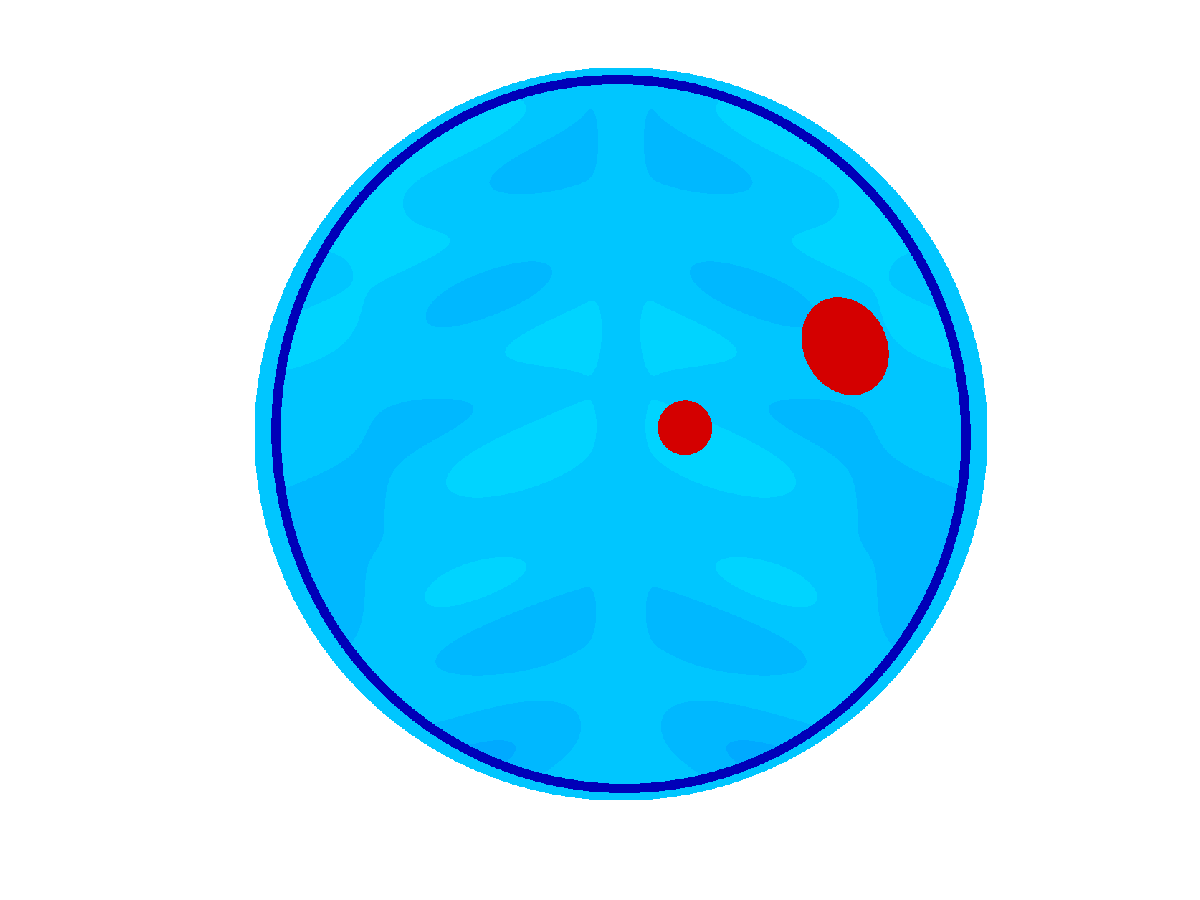}}
	\end{picture}    
	\caption{The testing dataset for the Neural Networks.  Top left: an example of the testing data set with circular inclusion. Top right: an example of testing dataset with elliptic inclusion. Bottom left: an example of the testing data set that is irregular shaped inclusion. Bottom right: an example of the testing data set with multiple inclusions. As with the training data, in the testing data set, the $2$-Dimensional cross section of the brain is modeled by a unit disk. There are three regions in the brain, excluding the inclusion. The scalp with uniform background conductivity $1$, the elliptical skull with homogeneous conductivity and the interior of the brain with inhomogeneous conductivity. The width and conductivity of the skull, the conductivity of the inner brain and the area, location and conductivity of the inclusion are all sampled randomly from a uniform distribution in the range listed in the Table \ref{table:sim_data} for the testing data sets of circular, elliptic and irregular inclusions. The testing data of multiple inclusions contain different skull thickness and orientation than that of training data. The inclusion for the testing data set is located in the right side of the disk as with training data set.}     
	\label{fig:testing_data}
\end{figure}

For the case of circular, elliptic and irregular inclusions, all other parameters in the testing data set remains the same as that of training data set except for the shape of the inclusion. However, in the case of multiple inclusions, we consider two inclusions of different shape; one of the inclusion is circular in shape and another one is elliptical in shape. In the case of multiple inclusions, we change the width of the skull to be in the range of $[0.03,0.05]$, thus making it different from the training set. The testing data of the multiple inclusions also contain different orientation than that of training data. This was achieved by rotating the skull by $5^{\degree}$ from the $y$-axis. Figure \ref{fig:testing_data} represents the four different inclusions of the testing data set. 

For each of the simulated $2$-Dimensional cross section of the head model, we compute EIT data, ${\bf{L}}_{\sigma}$ as described in Section \ref{sec:DN}. We add relative noise $\delta = 10^{-3}$ and $\delta = 10^{-2}$ and compute the noisy DN matrix ${\bf{L}}_{\sigma}^{\delta}$. We also compute the VHED function, $\widehat{\omega}(z,t,e^{i \phi})$ from ${\bf{L}}_{\sigma}$ and ${\bf{L}}_{\sigma}^{\delta}$. 

 We list the two different types of training pairs used to train our networks separately below.
\begin{itemize}
    \item[(i)] DN matrix training data, comprising of training pair $\{{\bf{L}}_{\sigma_j},y_j \}$  
    \item[(ii)] VHED training data, comprising of training pair $\{\widehat{\omega}(z,t,1),y_j \}$ 
\end{itemize}
We describe the training pairs used for FCNN and CNN in detail in  the following sections \ref{sec:FCNN} and \ref{sec:CNN}, respectively.

\begin{table}
	\begin{center}
		\begin{tabular}{|p{6.0cm}|p{2.0cm}|}

			\hline
			Width of the skull                   &  $[0.03, 0.04]$ 	\\
			\hline
			Conductivity of the skull            &  $[0.1, 0.2]$     \\
			\hline
			Conductivity of the brain            &  $[0.8, 1.1]$     \\
			\hline 
			Radius of the stroke                 &   $[0.1, 0.15] $	 \\
			\hline
			Conductivity of Ischemic stroke      & $[0.7, 0.8]$      \\
			\hline
			Conductivity of Hemorrhagic stroke   & $[3, 4]$          \\
			\hline   	  		
		\end{tabular}
	\end{center}
	\caption{The ranges for the conductivity distribution for the various parts of the simulated two-dimensional cross section of the brain used in generating the training and testing data set.}
	\label{table:sim_data}		
\end{table}

\subsection{Fully Connected Neural Network (FCNN)}\label{sec:FCNN}

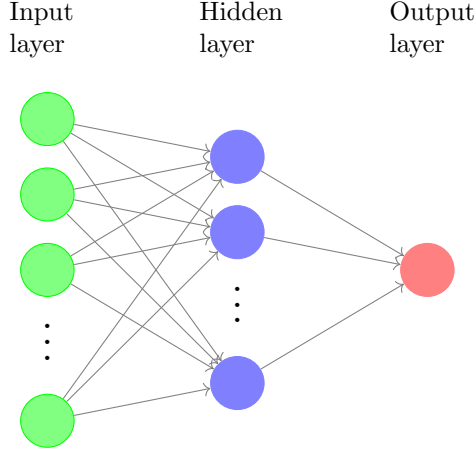
\begin{figure}
\centering 	
\begin{tikzpicture}

\node[circle, draw=green, fill=green!50,minimum size=20pt,inner sep=0pt] (G1)  at (0,5)  {};
\node[circle, draw=green, fill=green!50,minimum size=20pt,inner sep=0pt] (G2)  at (0,4)  {};
\node[circle, draw=green, fill=green!50,minimum size=20pt,inner sep=0pt] (G3)  at (0,3)  {};
\node[circle, draw=green, fill=green!50,minimum size=20pt,inner sep=0pt] (G4)  at (0,1)  {};
\node at (0,2.2) {\scalebox{1.5}{$\vdots$}};

\node[circle, draw=blue!50, fill=blue!50,minimum size=20pt,inner sep=0pt] (B1)  at (2.5,4.5)  {};
\node[circle, draw=blue!50, fill=blue!50,minimum size=20pt,inner sep=0pt] (B2)  at (2.5,3.5)  {};
\node[circle, draw=blue!50, fill=blue!50,minimum size=20pt,inner sep=0pt] (B3)  at (2.5,1.5)  {};
\node at (2.5,2.7) {\scalebox{1.5}{$\vdots$}};		

\node[circle, draw=red!50, fill=red!50,minimum size=20pt,inner sep=0pt] (R)  at (5,3)  {};

\draw[gray,->]  (G1)--(B1);
\draw[gray,->]  (G1)--(B2);
\draw[gray,->]  (G1)--(B3);

\draw[gray,->]  (G2)--(B1); 
\draw[gray,->]  (G2)--(B2); 
\draw[gray,->]  (G2)--(B3);

\draw[gray,->]  (G3)--(B1); 
\draw[gray,->]  (G3)--(B2); 
\draw[gray,->]  (G3)--(B3);

\draw[gray,->]  (G4)--(B1); 
\draw[gray,->]  (G4)--(B2); 
\draw[gray,->]  (G4)--(B3);

\draw[gray,->]  (B1)--(R); 
\draw[gray,->]  (B2)--(R); 
\draw[gray,->]  (B3)--(R);   

\node[above, text width=1cm] at (0  ,5.7) {Input layer};

\node[above, text width=1cm] at (2.5,5.7) {Hidden layer};

\node[above, text width=1cm] at (5  ,5.7) {Output layer};
\end{tikzpicture}
\caption{Illustration of the FCNN architecture considered in this study. The network consists of one input layer with 1024 nodes, one hidden layer with 30 nodes and one output layer with 1 node.} \label{fig:FCNN}
\end{figure}

The FCNN considered in this study is a shallow network with an architecture that consists of one input layer, one hidden layer and one output layer, see Figure~\ref{fig:FCNN}.

We recall that the DN matrix ${\bf{L}}_{\sigma}$ is a $33 \times 33$ real-valued matrix. We remove the first row and first column corresponding to the constant basis function $\phi_0$. The resulting matrix is represented as a real-valued column vector in $\mathbb{R}^{1024}$ which is used as the input for the DN matrix training pair.

For the second training pair, the input of VHED function, $ \widehat{\omega}(z,t,1) $ is computed at $z=\pm 1$. The pseudo-time $t \in [-2,2]$ was discretized over $256$ points, thus obtaining $\widehat{\omega}(z,t,1) \in \mathbb{C}^{256}$. Then, the real and imaginary parts of the VHED profiles at $z=\pm 1$ were concatenated as real-valued column vector in $\mathbb{R}^{1024}$. 

The hidden layer of the FCNN has $30$ neurons and the activation function is a sigmoid function. We consider different number of neurons for the DN matrix and VHED training pair and note that $30$ neurons in the hidden layer gives the best results.  

Finally the output layer consists of one neuron. The output is a scalar denoted by $y_j$:  $y_j$ being $0$ represents the low conductivity inclusion and $y_j$ being $1$ represents the high conductivity inclusion. 

Mathematically, FCNN can be represented as function $\mathcal{F}_{\theta}:\mathbb{R}^{1024}\to \mathbb{R}$. If we denote input by $x^{(0)}\in \mathbb{R}^{1024}$, the weight matrix at the first layer by $W^{1}_{\theta} \in \mathbb{R}^{30 \times 1024}$ and the vector of biases of this first layer by $b^{1}_{\theta} \in \mathbb{R}^{30}$, then the input of the second layer is given by
$$x^{(1)}=f(W_{\theta}^1x^{(0)}+b_{\theta}^{1}),$$
where $f$ is the sigmoid activation function defined by $f(t)=(1+e^{-t})^{-1}$. Note that $f$ is applied to every element of the vector. 
Furthermore, let $W_{\theta}^{2} \in \mathbb{R}^{1 \times 30}$ and  $b_{\theta}^{2} \in \mathbb{R}$ be the weight and the bias of the second layer. Then, the output of the FCNN is given by   
$$y=f(W_{\theta}^2x^{(1)}+b_{\theta}^{2})=f(W_{\theta}^2f(W_{\theta}^1x^{(0)}+b_{\theta}^{1})+b_{\theta}^2)$$
Thus, the FCNN can be can be represented in the following compact form
$$y=\mathcal{F}_{\theta}(x^{(0)})$$
where $\theta \in \mathbb{R}^{30781}$ is a vector containing all the network parameters (weights and biases).
The network $\mathcal{F}_{\theta}$ is then trained to find the optimal parameters.
This is done by minimizing the binary cross entropy loss function,
\begin{equation}\label{bi_cross_entropy}
L (\theta) = \sum_{j}-y_j \text{log}(y^p_j) - (1-y_j) \text{log}(1-y^p_j),
\end{equation}
where $y^p_j= \mathcal{F}_{\theta}(x_j)$ is the predicted value of outcome, $y_j$ is the true value of the outcome and the sum is over all training inputs. We use the scaled conjugated gradient algorithm to minimize the loss function. 

The results of the FCNN used for the classification of stroke using both DN matrix and VHED training pairs separately are shown in Section~\ref{section:FCNN}. 

\subsection{Convolutional Neural Network (CNN)} \label{sec:CNN}

Mathematically, the CNN can be defined with a function $f_{\theta}: \mathbb{R}^{d_0} \rightarrow \mathbb{R}^{d_{L}}$ with depth $L$ and $\theta = \{A_{\theta}^{l} , b_{\theta}^{l} \}$, a set of weights $A_{\theta}^{l}$ and biases $b_{\theta}^{l}$, 
\begin{equation}\label{CNN_fun}
     f_{\theta}(x) = a_{L},
\end{equation}
Here, the input to the CNN is taken to be  $a_0$. Let the index $ l = 0,1,2,.. L$ denote the number of layers in CNN. The output of the layer $l$, $a_{l} \in \mathbb{R}^{d_l}$ is a tensor given by
\begin{equation}\label{layers_CNN}
    a_{l} =  \phi_{l}[A^{l}_{\theta} a_{l-1} + b^{l}_{\theta}],
\end{equation}
Here $d_l$ represents the width of the layer $l$ and $\phi_l: \mathbb{R}^{d_l} \rightarrow \mathbb{R}^{d_l}$ are the activation functions that apply a scalar function to each of the component of the output tensor $a_l$. 

The output of a $2$ dimensional convolution applied on the $l$ layer is given by
\begin{equation}\label{conv_kernel}
 {a_{l}}(i,j) = b_{\theta}^{l} + \sum_{m} \sum_{n} A_{\theta}^{l}(m,n) a_{l-1}(i-m,j-m)
\end{equation}
Here, the weight $A_{\theta}^{l}$ is a convolution kernel of size $ m \times n$. 

A batch normalization is applied on the output of the Convolution layer. Batch normalization stabilizes the network parameters by normalizing the output of convolution layer. A brief description of the batch normalization applied on the $l$-th Convolution layer is as follows:
Let the output of the $l$-th convolution layer, denoted by  $a_{l}$, be divided into mini batches of size $m$ given by $\mathcal{B} = \{{a}_{l}^1, {a}_{l}^2, \cdots , {a}_{l}^{m} \}$ . We compute the mean of $\mathcal{B}$ as 
\begin{equation*}
\mu_{\mathcal{B}} = \frac{1}{m} \sum_{n =1}^{m} {a}_{l}^{n},
\end{equation*}
and variance of $\mathcal{B}$ 
\begin{equation*}
    \sigma_{\mathcal{B}}^2 = \frac{1}{m} \sum_{n = 1}^{m} ({a}_{l}^{n} - \mu_{\mathcal{B}})^{2}.
\end{equation*}
Normalizing over the entire batch $\mathcal{B}$, 
\begin{equation*}
    \widehat{a}_{l}^{n} = \frac{{a}_{l}^{n} - \mu_{\mathcal{B}}}{\sqrt{\sigma_{\mathcal{B}}^2 +\epsilon}} \qquad n = 1,2,\cdots m .
\end{equation*}
We then scale and shift to get the output $a_{l}$ as
\begin{equation*}
    a_{l}^{n} = \gamma \widehat{a}_{l}^n +\beta  \qquad n = 1,2, \cdots m.
\end{equation*}
where $\gamma$ and $\beta$ are parameters to be learned.

A max pooling applied on layer $l$ results in the output $a_l$ that is reduced in dimension. We consider the matrix $a_{l-1}$ of dimension $p \times s$ from the output of layer of  $l-1$ and apply a $m$ row partition  and $n$ column partition to it. That is
\begin{equation*}
    a_{l-1} = \left[ \begin{array}{c c c }
    A_{11}  & \cdots  &A_{1r} \\
    A_{21} & \cdots  &A_{2r} \\
    \vdots & \ddots & \vdots \\
     A_{q1} & \cdots  & A_{qr} \end{array}\right]
\end{equation*}
 where $q = \lfloor \frac{p}{m} \rfloor$ and $r = \lfloor \frac{s}{n} \rfloor$, $p \geq m$, $q \geq n$. 

In each of the matrices $A_{ij}(i',j')$ we choose the maximum element as the element of the output matrix given by, 
\begin{equation}\label{max_pool}
    a_{l}(i',j') = \text{max}\{A_{ij}(i',j')\}
\end{equation}
where $1 \leq q$, $1 \leq j \leq r$, $1 \leq i' \leq m$ and $1 \leq j' \leq n$.

We use two kinds of activation functions in the architecture of CNN. For the output of the $l$-th convolution layer, $a_{l}(i,j)$, we use a leaky rectified linear function given by
\begin{equation}\label{leaky_relu}
    \phi_{l}(a_{l}(i,j)) = \left\{ \begin{array}{c}
     0.001a_{l}(i,j) \qquad \text{for} \qquad a_{l}(i,j) \leq 0, \\
     a_{l}(i,j) \qquad \text{for} \qquad a_{l}(i,j)>0 \end{array} \right.
\end{equation}
 
We use the function softmax as an activation function for the output vector of fully connected layer $a_l \in \mathbb{R}^{d_l}$, given by
\begin{equation}\label{soft_max}
    \phi_{l}(a_{l})_{i=1}^{d_l} = \frac{e^{a_{l}^i}}{\sum\limits_{j = 1}^{d_l} e^{a_{l}^{j}}}.  
\end{equation}

We first train the network with the DN matrix training data and then train a CNN with a different architecture with the VHED training data. For both DN matrix and VHED training data, the output $y_j$ is a categorical vector $\{0,1\}$ with $\{0\}$ representing the low conductivity inclusion, that is ischemic stroke and $\{1\}$ representing the high conductivity inclusion, that is hemorrhagic stroke.

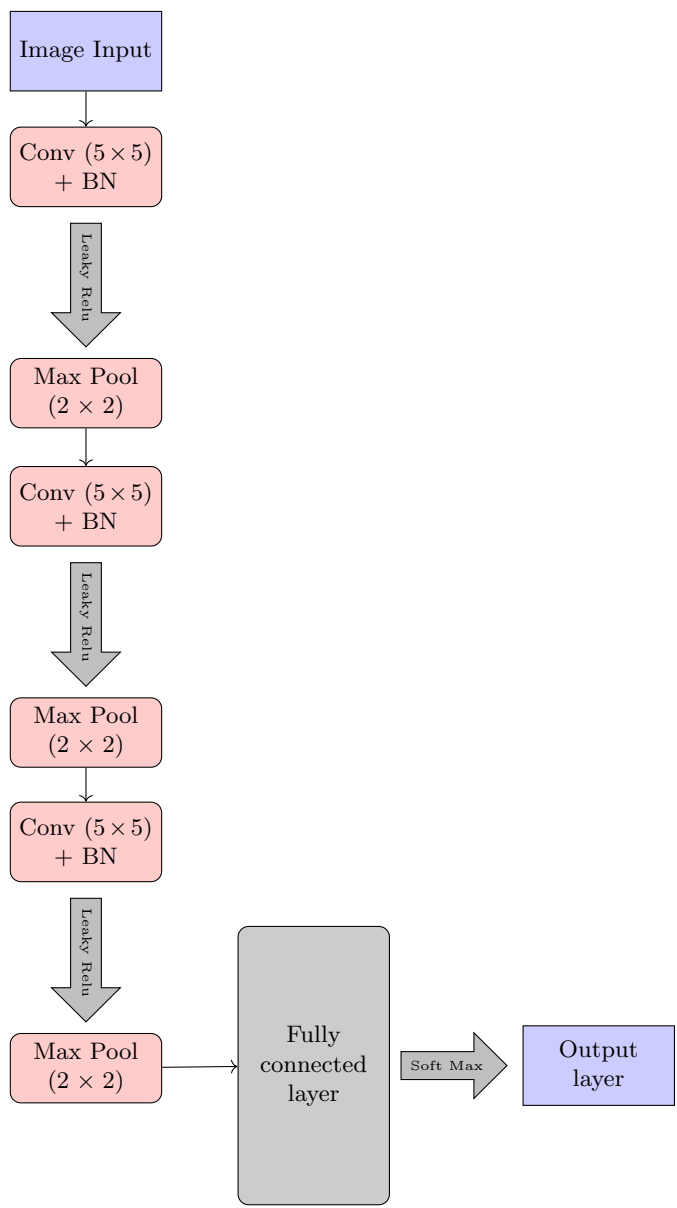
\begin{figure}
\centering
\begin{tikzpicture}

     \node (II) [II] {Image Input};
    \path (II.south)+ (0,-1) node (CBN)[CBN] {Conv ($5\times5$) + BN};
     \path (II.south)+ (0,-4) node (maxpool)[maxpool] {Max Pool ($2\times 2$) };
 \path (II.south)+ (0,-5.5) node (CBN_1)[CBN_1] {Conv ($5\times5$) + BN};
    \path (II.south)+ (0,-8.5) node (maxpool_1)[maxpool_1] {Max Pool ($2\times 2$)};
    \path (II.south)+ (0,-9.95) node (CBN_2)[CBN_2] {Conv ($5\times5$) + BN};
   \path (II.south)+ (0,-12.95) node (maxpool_2)[maxpool_2] {Max Pool ($2\times 2$)};
    \path (II.west)+ (4,-13.45) node (FC_1)[FC_1] {Fully connected layer};
    \path (II.west)+ (7.75,-13.45) node (OP)[OP] {Output layer} ;

  \draw [->] (II) (II)--(CBN) ; 
   \draw [->] (II) (maxpool)--(CBN_1) ;
  \draw[->](II) (maxpool_1)--(CBN_2) ;
   \draw[->](II) (maxpool_2)--(FC_1) ;
 
   \node[draw,single arrow, black ,fill=gray!50, font = \tiny ,rotate = -90] at (0,-3,0) {Leaky Relu};
   \node[draw,single arrow, black ,fill=gray!50, font = \tiny,rotate = -90] at (0,-7.5,0) {Leaky Relu};
    \node[draw,single arrow, black ,fill=gray!50, font = \tiny,rotate = -90] at (0,-11.95,0) {Leaky Relu};
  \node[draw,single arrow, black ,fill=gray!50, font = \tiny] at (4.75,-13.45,0) {Soft Max};

\end{tikzpicture}
\caption{The CNN architecture used in the VHED training pair. This architecture consists of $9$ layers. The input layer, followed by a convolution layer with $15$ filters of the size $5 \times 5$. Next follows a Maximum pooling layer with a stride of $2 \times 2$ via Leaky Rectified Unit function. Then again a Convolution layer follows with $20$ convolution filters of size $5 \times 5$. This is followed by a maximum pooling layer with a stride $2 \times 2$ . This is followed by a convolution layer with $25$ filters of size $5 \times 5$ along with batch normalization. Again a max pooling layer is followed with an activation function of leaky rectified unit. The outputs of all the convolution layer undergoes batch normalization with $60$ mini-batches. The eighth layer is fully connected layer which is connected to the output via the softmax activation function.}
\label{fig:CNN_VHED}
\end{figure}

\noindent
\begin{enumerate}[(i)]
\item CNN architecture for the DN matrix training pair. 

The CNN architecture for the DN matrix training has four layers. The input layer given by  $a_0 = {\bf{L}}_{\sigma}$. Since ${\bf{L}}_{\sigma} \in \mathbb{R}^{33 \times 33}$, we have $a_0 \in \mathbb{R}^{33 \times 33 \times 1}$. The third dimension in the tensor, $a_0$ corresponds to the number of channels.

The second layer consists of a $2$-dimensional convolution layer with $10$ convolution filters of the size $3\times3$. The output of this layer given by \eqref{conv_kernel} is followed by a batch normalization layer with $10$ mini batches. The Leaky rectified linear activation function maps the output to a fully connected layer. The Softmax activation function maps the output of the fully connected layer to the final Output layer.

\begin{figure}
\centering
\begin{tikzpicture}
    
    \node (II) [II] {Image Input};
    \path (II.south)+ (0,-1) node (CBN)[CBN] {Conv  ($3\times3$)  + BN};
    
     \path (II.south)+ (0,-4.25) node (FC)[FC] {Fully Connected Layer};
     
      \path (II.south)+ (0,-7.05) node (OP)[OP] {Output layer} ;

   \draw [->] (II) (II)--(CBN) ; 
   \node[draw,single arrow, black ,fill=gray!50, font = \tiny, rotate = -90] at (0,-3,0,0) {Leaky ReLU};
   \node[draw,single arrow, black ,fill=gray!50, font = \tiny,rotate = -90] at (0,-6.15,0) {Softmax};
\end{tikzpicture}
\caption{The CNN architecture used in the classification of the DN matrix training pair. This architecture consists of $4$ layers. The input layer, followed by a convolution layer with $10$ filters of the size $3 \times 3$ along with batch normalization on $10$ mini-batches and a fully connected layer and output layer. We apply the activation of leaky rectified function given by \eqref{leaky_relu} to the output of Convolution and batch normalization layer and activation of Softmax function to the output of fully connected layer. }
\label{fig:CNN_DN}
\end{figure}
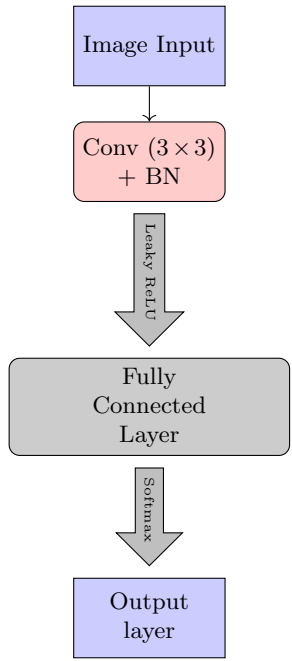

\item CNN architecture for the VHED training pair.

 The input of the VHED training pair is given by $\widehat{\omega}(z,t,1)$. It is computed on $z =64$ points along the boundary, $\partial \Omega$. We discretize the pseudotime in the interval $t \in [-2,2]$ over $256$ points. We use both real and imaginary parts of the VHED function. Due to the availability of more features in the form of VHED functions, $\widehat{\omega}(z,t,1)$, we choose a different CNN architecture optimized for this particular input. The CNN architecture now consists of $9$ layers. 

The input layer is given by $a_0 = \widehat{\omega}(z,t,1) $. We note that $a_0 \in \mathbb{R}^{128 \times 256 \times1}$. The dimension of $128 \times 256$ accounts for $64 \times 256$ real part and $64 \times 256 $ imaginary parts of the complex valued VHED functions. The third dimension of the tensor $a_0$ corresponds to the number of channels used.

The second layer in the CNN is a $2$-D convolution layer with $15$ kernels of he size $5 \times 5$; the output of which is given by \eqref{conv_kernel}. This is followed by batch normalization with $60$ mini-batches. The leaky rectified linear activation function maps the output of this layer to the third layer of maximum pooling with a partition of $2 \times 2$. The output of the maximum pooling layer is given by \eqref{max_pool}.

The fourth layer is CNN is a $2$ dimensional convolution layer with $20$ kernels of size $5 \times 5$. This output given by \eqref{conv_kernel} undergoes batch normalization with $60$ mini batches. The leaky rectified linear activation function maps this output to a maximum pooling with partition of $2 \times 2$; the output of which is given by \eqref{max_pool}.

The sixth layer is again a $2$ dimensional convolution layer with $25$ convolution kernels of size $5 \times 5$. The output given by \eqref{conv_kernel} undergoes batch normalization with $60$ mini batches. The leaky rectified linear activation function maps this output to a maximum pooling layer with a partition of $2 \times 2$.

The eighth layer is a fully connected layer. The Softmax activation function maps the output of this layer to the final Output layer of the CNN. Figure \ref{fig:CNN_VHED} represents the different layers in the CNN architecture used for the VHED training pair.
\end{enumerate}

We note that the size of the inputs to the CNN differs from that of FCNN. The convolution kernels in CNN are capable of handling network parameters of higher dimension without much cost to the computational efficiency. Thus, we choose to include all possible features available as the input of CNN which results in high dimensional network parameters, $\theta$. We note that this is not the only way of representing the features.

We use {\textit{$k$-fold}} cross-validation method, where $k = 5$ to train the CNN networks for DN matrix and VHED training pair. Both the CNNs $f_{\theta}$ are trained to find the optimal parameters $\theta$ by minimizing the binary cross entropy loss function
\begin{equation}
     L(\theta) = \sum_{j=1}^{N} -y^t_j \text{log}(y^p_j) - (1-y^t_j) \text{log}(1-y^p_j).
\end{equation}
using the ADAM algorithm \cite{article}. Here $y_j^p$ is the predicted value of the outcome, $y_j^t$ is the true value of the outcome and the summation is over $N$ training inputs. The results of the CNNs used for the classification of stroke using both DN matrix and VHED training pairs separately are shown in Section \ref{section:CNN}. 

\clearpage

\section{Results}\label{sec:results}

In this section we present the results of classification of the inclusion into low or high conductivity with the the DN matrix training pair and the VHED training pair using the networks, FCNN and CNN.

To train the networks, FCNN and CNN, we use dataset with circular inclusions. We test the trained networks with three different datasets: (a) Circular inclusions, (b) Elliptical inclusions (c) Irregular shaped inclusions and (d) Multiple inclusions. The elliptic inclusions and the irregular shaped inclusions differ in shape from the training dataset. The Multiple inclusions dataset differ from from the training set in multiple ways. The dataset contains more number of inclusions than the training data. The width of the skull is increased compared to the training dataset. The skull contains different orientation compared to the training dataset.

We recall that the training of networks, FCNN and CNN, is done with two training pairs: The DN matrix training pair and and the VHED training pair.
We add relative noise of $\delta = 10^{-3}$ and $\delta = 10^{-2}$ to the DN matrix, ${\bf{L}}_{\sigma}$ and compute the VHED functions that correspond to the noisy DN matrix. The relative noise is added to both training dataset and the testing dataset. The details of computation of the DN matrix and VHED function are described in Section \ref{section:Data}. 

We perform three separate trainings of networks, with varying levels of noise. 

First we train the networks with no added noise in the simulation. The intrinsic relative numerical error of the FEM computations on the DN matrix is approximately $10^{-5}$. So both the training data and the test data have no added noise, but still suffer from slight degradation whose amplitude we call $10^{-5}$. All of this is done for both DN matrix training pair and and the VHED training pair.

Second, we move on to work with simulated noise. We add noise with relative amplitude of  $10^{-3}$ to the DN maps, and from there we calculate the VHED profiles. Then the networks are trained with training data suffering from slight degradation. Both the training data and the testing data suffer from slight degradation whose amplitude is $10^{-3}$.

Third, we repeat the noisy training as above but adding simulated noise with relative amplitude of  $10^{-2}$ to the ND maps.


To measure the performance of networks, we use the metric of sensitivity, specificity and accuracy on the testing data set. \\[10pt]
{\bf Sensitivity} measures the proportion of actual positives that are correctly identified as such.\\[10pt]
{\bf Specificity} measures the proportion of actual negatives that are correctly identified as such. \\[10pt]
{\bf Accuracy} is the sum of true positives and true negatives over the total test population, that is, the fraction of correctly classified cases.\\[10pt]

Here, we recall that the outcome of the neural networks are either labeled $\{0\}$ indicating low conductivity inclusion or labeled $\{1\}$ indicating high conductivity inclusion. In this study positive outcome corresponds to the label $\{1\}$ and negative outcome corresponds to the label $\{0\}$.

\subsection{Results of FCNN}\label{section:FCNN}
 
 We report the performance metrics of trained FCNN that was tested on  four different data sets. We use $N=4000$ circular inclusion data set to train the FCNN.
 
We recall that (a) the DN matrix training pair  for FCNN is given by ${\bf{L}}_{\sigma} \in \mathbb{R}^{1024}$ and (b) the VHED training pair is given by $\widehat{\omega}(z,t,1) \in \mathbb{R} ^{1024}$. 

We train the FCNN with the DN matrix training pair and then train the same FCNN with other training pair of VHED functions. The architecture of the FCNN is described in Section \ref{sec:FCNN}.
 
The performance metrics reported for the FCNN is the average value computed over $20$ FCNN trainings. We shuffle the data randomly and choose the initial weights and biases randomly during each of the training. The performance metrics evaluated on three data sets are:

\begin{enumerate}[(i)]

\item Circular inclusions: We report the performance metrics of the trained FCNN that was tested on $4000$ circular inclusions in the Table \ref{table:case1FCNN}. We note that the testing data set was not used in training.

\item Elliptic inclusions: We report the performance metrics of the trained FCNN that was tested on $2000$ elliptic inclusions in the Table \ref{table:case2FCNN}.

\item Irregular shaped inclusions: We report the performance metrics of the trained FCNN that was tested on $1000$ irregular inclusions in the Table \ref{table:case3FCNN}.

Multiple inclusions: We report the performance metrics of the trained FCNN that was tested on $1000$ multiple inclusions in the Table \ref{table:case4FCNN}.   

\end{enumerate}

With all the four different inclusions tested the accuracy of the prediction is better with VHED training pair. The testing datasets of elliptic, irregular shape and multiple inclusions are significantly different datasets than the one used in training. The accuracy is better for the irregular inclusion dataset than it is for elliptic inclusion and multiple inclusion dataset for both training pairs. For the testing data sets of elliptic inclusions and irregular inclusions, the accuracy remains either the same or differs very little when using the VHED training pair computed from the noisy DN matrix, suggesting that noise has very little effect on the classification when using VHED training pair. For the testing dataset of multiple inclusions the accuracy improves slightly when using the VHED training pair computed from the noisy DN matrix. The phenomenon that adding of noise into the training data may improve the classification has been studied for example in \cite{Bishop-1995}. There, it is shown that adding noise to training data is equivalent to modifying the regularizer in the training process, and thus adding noise into the training data may reduce the overfitting and improve the accuracy of the classification.

We present in Table~\ref{table:FCNNStats} the mean and the standard deviation of the accuracy for the four different testing datasets. The standard deviation of the accuracy for the four different testing datasets is smaller for the VHED training pair, particularly when the VHED is computed from noisy DN matrix. 

\subsection{Results of CNN}\label{section:CNN}
In this section we report the performance metrics of trained CNNs that were tested on four different data sets. We use $N=8000$ circular inclusion data set to train the CNNs. 

We recall that (a) the DN matrix training pair for CNN is given  by ${\bf{L}}_{\sigma} = \mathbb{R}^{33 \times 33 \times 1}$ and (b) the VHED training pair is given by $\widehat{\omega}(z,t,1) \in \mathbb{R} ^{128 \times 256 \times 1}$. 

We train the CNN with DN matrix training pair and then train another CNN with VHED training pair. The two architectures of the CNNs used for the two different training pairs are described in Section \ref{sec:CNN}. 

We evaluate the performance of the CNNs on the four different testing datasets. We note that all the measures of the performance are the mean of the $5$-fold training of the CNNs.

\begin{enumerate}[(i)]

\item  Circular inclusions: We report the performance metrics of the trained CNNs that were tested on $2000$ circular inclusions in Table \ref{table:CNN_circ}.  The testing data set was not used in the training.   

\item Elliptic inclusions: We report the performance metrics of the trained CNNs that were tested on $2000$ elliptic inclusions in Table \ref{table:CNN_ell}. 

\item Irregular shaped inclusions: We report the performance metrics of the trained CNNs that were tested on $1000$ irregular inclusions in Table \ref{table:CNN_irr}.

\item Multiple inclusions: We report the performance metrics of the trained CNN that was tested on $1000$ multiple inclusions in the Table \ref{table:CNN_multiple}.

\end{enumerate}

As with the FCNN, with all the four different inclusion data sets tested, the accuracy of the prediction is better with VHED training pair. For the irregular and multiple inclusions testing dataset, the accuracy is significantly better with the VHED training pair compared to DN matrix training pair. With elliptic inclusion and multiple inclusion testing dataset, the accuracy is better with the VHED training pair, where the VHED functions are computed from the noisy DN matrix. Like with FCNN, adding noise to training data may reduce the overfitting and improve the accuracy of the classification \cite{Bishop-1995}. 

We present in Table \ref{table:Stats} the mean and the standard deviation of the accuracy of the trained CNNs for the four different testing datasets.

 \clearpage

\begin{table}
\begin{center}
\begin{tabular}{|p{2.5cm}||p{1.0cm}|p{1.0cm}|p{1.0cm}||p{1.0cm}|p{1.0cm}|p{1.0cm}|}
\hline 					

  \multicolumn{7}{|c|} {FCNN: Circular Inclusions} \\
\hline
 	   			& \multicolumn{3}{c||}{ DN matrix training pair}  & \multicolumn{3}{c|}{ VHED training pair} \\
 	   			\hline
 	   			Noise $\delta$   & $10^{-5}$   &  $10^{-3}$ &  $10^{-2}$ &  $10^{-5}$ & $10^{-3}$ & $10^{-2}$
 	   			\\			
 	   			\hline
 	   			Sensitivity           &  0.9996   &  0.9995    & 0.9956 
 	   			                      &    1      &  1         & 0.9992    \\	
 	   			\hline 
 	   			Specificity           &    1      &  1         &  0.9994
 	   			                      &    1      &  1         &  0.9997		\\	
 	   			\hline
 	   			Accuracy              &  0.9998   &  0.9997    &  0.9975
 	   			                      &    1      &  1         &  0.9994   	\\	
 	   			\hline   	  		
\end{tabular}
\end{center}
\caption{ The performance metrics of FCNN trained with the two different training pairs and evaluated on datasets consisting of $4000$ circular inclusions.}%
\label{table:case1FCNN}		
\end{table}

\begin{table}
\begin{center}
\begin{tabular}{|p{2.5cm}||p{1.0cm}|p{1.0cm}|p{1.0cm}||p{1.0cm}|p{1.0cm}|p{1.0cm}|}
  			\hline
  			
  \multicolumn{7}{|c|} {FCNN: Elliptic Inclusions} \\
\hline
  			& \multicolumn{3}{c||}{ DN matrix training pair}  & \multicolumn{3}{c|}{ VHED training pair} \\
  			\hline
  			Noise $\delta$    & $10^{-5}$   &  $10^{-3}$ &  $10^{-2}$ & $10^{-5}$  & $10^{-3}$ & $10^{-2}$
  			\\			
  			\hline
  			Sensitivity             & 0.8563    & 0.8560    & 0.8081 
  			                        & 1         & 0.9998    & 0.9651
  			\\		
  			\hline
  			Specificity             & 1         & 1         & 1 
  			                        & 1         & 1         & 1
  			\\	
  			\hline
  			Accuracy                & 0.9259    & 0.9250    & 0.9001 
  		                         	& 1         & 0.9999    & 0.9818 	
  			\\	
  			\hline   
\end{tabular}	
\end{center}
\caption{ The performance metrics of FCNN trained with the two different training pairs and evaluated on datasets consisting of $2000$ elliptic  inclusions.}%
\label{table:case2FCNN}		
\end{table}

\begin{table}
\begin{center}
\begin{tabular}{|p{2.5cm}||p{1.0cm}|p{1.0cm}|p{1.0cm}||p{1.0cm}|p{1.0cm}|p{1.0cm}|}
  			\hline
  			
  \multicolumn{7}{|c|} {FCNN: Irregular Inclusions} \\
\hline
  			& \multicolumn{3}{c||}{ DN matrix training pair}  & \multicolumn{3}{c|}{ VHED training pair} \\
  			\hline
  			Noise $\delta$    & $10^{-5}$   &  $10^{-3}$ &  $10^{-2}$ &  $10^{-5}$ & $10^{-3}$ & $10^{-2}$
  			\\			
  			\hline
  			Sensitivity             & 0.9760    & 0.9715    & 0.9627 
  			                        & 1         & 1         & 0.9969
  			\\		
  			\hline
  			Specificity             & 1         & 1         & 1 
  			                        & 1         & 1         & 1
  			\\	
  			\hline
  			Accuracy                & 0.9879    & 0.9855    & 0.9811 
  			                        & 1         & 1         & 0.9984 	
  			\\	
  			\hline   
 \end{tabular}	
 \end{center}
 \caption{ The performance metrics of FCNN trained with the two different training pairs and evaluated on datasets consisting of $1000$ irregular inclusions.}%
\label{table:case3FCNN}		
\end{table}

\begin{table}
    \begin{center}
        \begin{tabular}{|p{2.5cm}||p{1.0cm}|p{1.0cm}|p{1.0cm}||p{1.0cm}|p{1.0cm}|p{1.0cm}|}
			\hline
			
			\multicolumn{7}{|c|} {FCNN: Multiple Inclusions} \\
			\hline
			& \multicolumn{3}{c||}{ DN matrix training pair}  & \multicolumn{3}{c|}{ VHED training pair} \\
			\hline
			Noise $\delta$    & $10^{-5}$   &  $10^{-3}$ &  $10^{-2}$ &  $10^{-5}$ & $10^{-3}$ & $10^{-2}$
			\\			
			\hline
			Sensitivity       & 0.5614      & 0.5924     & 0.5203     & 0.9980     & 0.9999    & 0.9838
			\\		
			\hline
			Specificity       & 0.4960      & 0.5438     & 0.7160     & 0.9664     & 0.9948    & 0.9991
			\\	
			\hline
			Accuracy          & 0.5292      & 0.5673     & 0.6168     & 0.9824     & 0.9974    & 0.9914 	
			\\	
			\hline   
		\end{tabular}	
	\end{center}
	\caption{ The performance metrics of FCNN trained with the two different training pairs and evaluated on datasets consisting of $1000$  multiple inclusions.}%
	\label{table:case4FCNN}		
\end{table}

\begin{table}
	\begin{center}
		\begin{tabular}{|p{1.5cm}|p{1.5cm}|p{1.5cm}|p{1.5cm}|p{1.5cm}|p{1.5cm}|}
			
			
			\hline
			
			\multirow{1}{*}{Inclusion} &
			\multirow{1}{*}{Noise} &
			\multicolumn{2}{c|}{DN maps}&
			\multicolumn{2}{c|}{VHED profiles} \\
			\hline
			
			& & Mean & Standard deviation & Mean & Standard deviation\\
			
			\hline
			Circular
			& $10^{-5}$       & 0.9998 & 0.001  & 1      & 0     
			\\			
			\hline
			Circular	 & $10^{-3}$ & 0.9997 & 0.0001 & 1      & 0	
			\\		
			\hline
			Circular	 & $10^{-2}$ & 0.9975 & 0.0006  & 0.9994 & 0.0003
			\\	
			\hline
			\hline
			Elliptic & $10^{-5}$        & 0.9259 & 0.0001  &1       & 0
			\\	
			\hline   
			Elliptic & $10^{-3}$  & 0.9250 & 0.0085  &0.9999  &0.0002
			\\			
			\hline
			Elliptic & $10^{-2}$  & 0.9001 & 0.0133  &0.9818  &0.0058 
			\\
			\hline
			\hline
			Irregular & $10^{-5}$       & 0.9879 & 0.0035  & 1      & 0
			\\	
			\hline   
			Irregular & $10^{-3}$ & 0.9855 & 0.0027  & 1      & 0
			\\			
			\hline
			Irregular & $10^{-2}$ & 0.9811 & 0.0046  & 0.9984 & 0.0006  
			\\
			\hline
			\hline
			Multiple & $10^{-5}$  & 0.5299 & 0.0069  & 0.9824    & 0.0170
			\\	
			\hline   
			Multiple & $10^{-3}$  & 0.5673 & 0.0255  & 0.9974    & 0.0041
			\\			
			\hline
			Multiple & $10^{-2}$  & 0.6168 & 0.0547  & 0.9914    & 0.0060  
			\\
			\hline
		\end{tabular}	
	\end{center}
	\caption{Mean and Standard deviation of the accuracy of the FCNN trained with two training pairs and tested on four different inclusions.}%
	\label{table:FCNNStats}		
\end{table}

\clearpage
\begin{table}
  \begin{center}
	\begin{tabular}{|p{2.5cm}||p{1.0cm}|p{1.0cm}|p{1.0cm}||p{1.0cm}|p{1.0cm}|p{1.0cm}|}
	\hline

  \multicolumn{7}{|c|} {CNN: Circular Inclusions} \\
\hline
		
		& \multicolumn{3}{c||}{ DN matrix training pair}  & \multicolumn{3}{c|}{ VHED training pair} \\
		\hline
		Noise $\delta$    & $10^{-5}$  &  $10^{-3}$ &  $10^{-2}$ &  $10^{-5}$ & $10^{-3}$ & $10^{-2}$
		\\			
		\hline
		Sensitivity                      & 0.9300    &  0.9389  & 0.9360
		                                 &1          & 1    & 1
		\\		
		\hline
		Specificity                      & 0.9989       &  0.9958      &  0.9918
		                                 & 1         &1          &1 
		\\	
		\hline
		Accuracy                         & 0.9753    &  0.9645  &  0.9611
		                                 &  1        &  1   &  1	
		\\	
		\hline   
	\end{tabular}	
\end{center}
		\caption{ The performance metrics of CNNs trained with two training pairs and evaluated on datasets consisting of $2000$ circular inclusions.}%
		\label{table:CNN_circ}		
\end{table}

\begin{table}
  \begin{center}
	\begin{tabular}{|p{2.5cm}||p{1.0cm}|p{1.0cm}|p{1.0cm}||p{1.0cm}|p{1.0cm}|p{1.0cm}|}
\hline	
  \multicolumn{7}{|c|} {CNN: Elliptic Inclusions} \\

		\hline
		& \multicolumn{3}{c||}{ DN matrix training pair}  & \multicolumn{3}{c|}{ VHED training pair} \\
		\hline
		Noise $\delta$    & $10^{-5}$   &  $10^{-3}$ &  $10^{-2}$ &  $10^{-5}$  & $10^{-3}$ & $10^{-2}$
		\\			
		\hline
		Sensitivity                      &  0.8913   &  0.8913   & 0.9199
		                                 & 0.9584        &  0.9855    & 0.9855
		\\		
		\hline
		Specificity                      &1     &   1     &  0.8356
		                                 &0.9390         & 0.9416         & 0.9416
		\\	
		\hline
		Accuracy                         & 0.9415  &  0.9415 &  0.8708
		                                 &  0.9600       &  0.9615   &  0.9615	
		\\	
		\hline   
	\end{tabular}	
\end{center}
		\caption{ The performance metrics of CNNs trained with two training pairs and evaluated on datasets consisting of $2000$ elliptical inclusions.}%
		\label{table:CNN_ell}		
\end{table}

\begin{table}
  \begin{center}
	\begin{tabular}{|p{2.5cm}||p{1.0cm}|p{1.0cm}|p{1.0cm}||p{1.0cm}|p{1.0cm}|p{1.0cm}|}
	\hline
 \multicolumn{7}{|c|} {CNN: Irregular Inclusions} \\
		\hline
		& \multicolumn{3}{c||}{ DN matrix training pair}  & \multicolumn{3}{c|}{ VHED training pair} \\
		\hline
		Noise $\delta$    & $10^{-5}$   &  $10^{-3}$ &  $10^{-2}$ &  $10^{-5}$ & $10^{-3}$ & $10^{-2}$
		\\			
		\hline
		Sensitivity                      & 1    &   0.7382  &  0.6787
		                                 &  0.9389     &   1   &  0.9360
		\\		
		\hline
		Specificity                      &  0.6552   &  1      &  1
		                                 &   0.9958     &   0.9306      &  0.9918
		\\	
		\hline
		Accuracy                         &  0.8252    & 0.8251   &  0.7667 
		                                 & 0.9645        &  0.9622   &  0.9611	
		\\	
		\hline   
	\end{tabular}	
\end{center}
		\caption{ The performance metrics of CNNs trained with two training pairs and evaluated on datasets consisting of $1000$ irregular inclusions.}%
		\label{table:CNN_irr}		
\end{table}

\begin{table}
    \begin{center}
        \begin{tabular}{|p{2.5cm}||p{1.0cm}|p{1.0cm}|p{1.0cm}||p{1.0cm}|p{1.0cm}|p{1.0cm}|}
			\hline
			
			\multicolumn{7}{|c|} {CNN: Multiple Inclusions } \\
			\hline
			& \multicolumn{3}{c||}{ DN matrix training pair}  & \multicolumn{3}{c|}{ VHED training pair} \\
			\hline
			Noise $\delta$    & $10^{-5}$   &  $10^{-3}$ &  $10^{-2}$ &  $10^{-5}$ & $10^{-3}$ & $10^{-2}$
			\\			
			\hline
			Sensitivity       & 0.4951     & 0.4955     & 0.4980     & 1     & 1    & 1
			\\		
			\hline
			Specificity       & 1      & 1     & 1     & 0.6522     & 0.6911    & 0.6911
			\\	
			\hline
			Accuracy          & 0.4951      & 0.4980     & 0.4980     & 0.7296     & 0.7734    & 0.7734 	
			\\	
			\hline   
		\end{tabular}	
	\end{center}
	\caption{ The performance metrics of CNN trained with the two different training pairs and evaluated on datasets consisting of $1000$  multiple inclusions.}%
	\label{table:CNN_multiple}		
\end{table}

\begin{table}
  \begin{center}
	\begin{tabular}{|p{1.5cm}|p{1.5cm}|p{1.5cm}|p{1.5cm}|p{1.5cm}|p{1.5cm}|}

	 
	\hline
	
	\multirow{1}{*}{Inclusion} &
	\multirow{1}{*}{Noise} &
	\multicolumn{2}{c|}{DN Maps}&
	\multicolumn{2}{c|}{VHED} \\
	\hline
	
	& & Mean & Standard deviation & Mean & Standard deviation\\

		\hline
Circular  & $10^{-5}$ & 0.9753 & 0.0125 & 1 &0
		\\			
		\hline
Circular  & $10^{-3}$ & 0.9645 &0.0166&1&0	
		\\		
		\hline
Circular  & $10^{-2}$ & 0.9611 & 0.0192 & 1 & 0
		\\	
		\hline
		\hline
Elliptic & $10^{-5}$ & 0.9415 & 0.0567 &0.9600 &0.0431
		\\	
		\hline   
Elliptic & $10^{-3}$ & 0.9415 & 0.1945 &0.9615 &0.0406
		\\			
	\hline
Elliptic & $10^{-2}$ & 0.8708 & 0.2085 &0.9615 &0.0405 \\
\hline
\hline
Irregular & $10^{-5}$ & 0.8252 & 0.0636 & 0.9645  & 0.0165
		\\	
		\hline   
Irregular & $10^{-3}$ & 0.8251 & 0.0635 & 0.9622 & 0.0169
		\\			
	\hline
Irregular & $10^{-2}$ & 0.7667 & 0.0534 & 0.9611 & 0.0192  \\
\hline
\hline
Multiple & $10^{-5}$ & 0.4964 & 0.0076 & 0.7296  & 0.0737 \\
\hline
Multiple & $10^{-3}$ & 0.4980 & 0.0082 & 0.7734  & 0.0922 \\
\hline
Multiple & $10^{-2}$ & 0.4980 & 0.0082 & 0.7734  & 0.0922 \\
\hline
	\end{tabular}	
\end{center}
		\caption{ A table of Mean accuracy  and Standard Deviation of the accuracy of the two different CNNs that were trained with two training pairs and tested on four different datasets.}%
		\label{table:Stats}		
\end{table}

\clearpage

\section{Discussion} \label{sec:Conclusions}
 
The numerical results from our study shows that the classification is  better from the VHED training pair for both neural networks, FCNN and CNN, for all the four different inclusions tested. 

For the FCNN, the results show good performance with both training pairs, DN matrix and VHED functions, in all the different testing data sets with exception of the test set having multiple inclusions where the network trained with  DN matrix shows a significant decrease in the accuaracy of classification.
In all four testing data sets the performance metrics are better for the VHED training pair. For the FCNN, in all the four testing datasets, the advantage of using VHED training pair is clearly evident with noisy training pairs. Additionally, the standard deviation is lower for all the four testing datasets, when the FCNN was trained with VHED functions, suggesting a more reliable accuracy for the classification. The computation of the VHED function $\widehat{\omega}(z,t,1)$ is noise robust due to the windowed Fourier Transform acting as a regularization. 

The CNNs, like FCNN, show better performance with VHED training pair for all the four testing cases. For the CNN, we note that the mean accuracy does not vary much  whether we train the network with DN matrix training pair or with the VHED training pair for the elliptic inclusion testing data set. For the CNN, we note that the mean accuracy of the classification with irregular shape is better than the mean accuracy of classification with either elliptic or multiple inclusions when using VHED training pair. The mean accuracy of the classification is significantly different for the irregular shaped inclusions and the multiple inclusions from the elliptic inclusions when using the DN matrix training pair.

We verify numerically with both FCNN and CNN that the VHED functions improve learning as noise-robust nonlinear input features. Furthermore, they lead to better interpretability of the classification due to the geometric properties explained in Section \ref{sec:VHED_th}. Therefore, the use of VHED functions as features can be seen as a  genuine Robust Grey-Box approach.

Our results also show that FCNN perform better than CNN even when using less training data and using fewer features of the VHED function.

Further research in this area is in the direction of including more realistic head shapes, smaller and irregular stroke-affected areas, and precise electrode modelling.

These complications lead to more severe challenges for the neural networks and may require the use of a more comprehensive collection of VHED profiles. This may lead to FCNNs with large training parameters that might become computationally ineffective. In such case the CNN approach may become necessary.

Also, it is important to test how the learning generalizes to data coming from three-dimensional targets. In lung imaging applications two-dimensional EIT reconstruction methods has been successfully applied to three-dimensional targets \cite{Mueller2018}, but in case of stroke imaging this may not be the case.

\section{Acknowledgements}

The work of Juan Pablo Agnelli was done during a research stay at University of Helsinki supported by the National Scientific and Technical Research Council of Argentina (CONICET). He also was partially supported by Secyt (UNC) grant 33620180100326CB. The work of Aynur Çöl was supported by a grant from the Scientific and Technological Research Council of Turkey (TUBITAK) to perform research at University of Helsinki.The work of Rashmi Murthy and Samuli Siltanen was supported in part by Jane and Aatos Erkko Foundation. The work of Matti Lassas, Rashmi Murthy and Samuli Siltanen was supported in part by the Center of Excellence in Inverse Modelling and Imaging,Academy of Finland, Decision number 312339. The work of Matteo Santacesaria is supported by Gruppo Nazionale per l'Analisi Matematica, la Probabilit\`a e le loro Applicazioni (GNAMPA) of the Istituto Nazionale di Alta Matematica (INdAM) Project 2019. Part of his work was carried at Machine Learning Genoa (MaLGa) center, Universit\`a di Genova (IT).

\bibliographystyle{plain}  
\bibliography{StrokeEIT.bib}

\begin{thebibliography}{10}

\bibitem{Alsaker19}
M.~Alsaker and J.L. Mueller.
\newblock Eit images of human inspiration and expiration using a {D}-bar method
  with spatial priors.
\newblock {\em Journal of the Applied Computational Electromagnetics Society
  (ACES)}, 34(2):325--330, 2019.

\bibitem{Alsaker18}
M.~Alsaker, J.L. Mueller, and R.~Murthy.
\newblock Dynamic optimized priors for {D}-bar reconstructions of human
  ventilation using electrical impedance tomography,.
\newblock {\em Journal of Computational and Applied Mathematics}, 362:276--294,
  2019.

\bibitem{arridge2019}
Simon Arridge, Peter Maass, Ozan Öktem, and Carola-Bibiane Schönlieb.
\newblock Solving inverse problems using data-driven models.
\newblock {\em Acta Numerica}, 28:1–174, 2019.

\bibitem{astala2006calderon}
Kari Astala and Lassi P{\"a}iv{\"a}rinta.
\newblock Calder{\'o}n's inverse conductivity problem in the plane.
\newblock {\em Annals of Mathematics}, pages 265--299, 2006.

\bibitem{bayford2012bioimpedance}
Richard Bayford and Andrew Tizzard.
\newblock Bioimpedance imaging: an overview of potential clinical applications.
\newblock {\em Analyst}, 137(20):4635--4643, 2012.

\bibitem{beals1984scattering}
Richard Beals and Ronald~R Coifman.
\newblock Scattering and inverse scattering for first order systems.
\newblock {\em Communications on Pure and Applied Mathematics}, 37(1):39--90,
  1984.

\bibitem{Bishop-1995}
Chris~M. Bishop.
\newblock Training with noise is equivalent to tikhonov regularization.
\newblock {\em Neural Computation}, 7, 1995.

\bibitem{boverman2016detection}
Gregory Boverman, Tzu-Jen Kao, Xin Wang, Jeffrey~M Ashe, David~M Davenport, and
  Bruce~C Amm.
\newblock Detection of small bleeds in the brain with electrical impedance
  tomography.
\newblock {\em Physiological measurement}, 37(6):727, 2016.

\bibitem{Calder'on1980}
A.-P. {C}alder{\'o}n.
\newblock On an inverse boundary value problem.
\newblock In {\em Seminar on {N}umerical {A}nalysis and its {A}pplications to
  {C}ontinuum {P}hysics ({R}io de {J}aneiro, 1980)}, pages 65--73. Soc. Brasil.
  Mat., Rio de Janeiro, 1980.

\bibitem{doshi2017towards}
Finale Doshi-Velez and Been Kim.
\newblock Towards a rigorous science of interpretable machine learning.
\newblock {\em arXiv preprint arXiv:1702.08608}, 2017.

\bibitem{edic1998iterative}
Peter~M Edic, David Isaacson, Gary~J Saulnier, Hemant Jain, and Jonathan~C
  Newell.
\newblock An iterative newton-raphson method to solve the inverse admittivity
  problem.
\newblock {\em IEEE transactions on biomedical engineering}, 45(7):899--908,
  1998.

\bibitem{faddeev2016increasing}
Lyudvig~Dmitrievich Faddeev.
\newblock Increasing solutions of the schr{\"o}dinger equation.
\newblock In {\em Fifty Years of Mathematical Physics: Selected Works of Ludwig
  Faddeev}, pages 34--36. World Scientific, 2016.

\bibitem{greenleaf2018propagation}
Allan Greenleaf, Matti Lassas, Matteo Santacesaria, Samuli Siltanen, and
  Gunther Uhlmann.
\newblock Propagation and recovery of singularities in the inverse conductivity
  problem.
\newblock {\em Analysis \& PDE}, 11(8):1901--1943, 2018.

\bibitem{hamilton2019beltrami}
Sarah~J Hamilton, Asko H{\"a}nninen, Andreas Hauptmann, and Ville Kolehmainen.
\newblock Beltrami-net: domain-independent deep d-bar learning for absolute
  imaging with electrical impedance tomography (a-eit).
\newblock {\em Physiological measurement}, 40(7):074002, 2019.

\bibitem{huhtanen2012numerical}
Marko Huhtanen and Allan Per{\"a}m{\"a}ki.
\newblock Numerical solution of the ℝ-linear beltrami equation.
\newblock {\em Mathematics of Computation}, 81(277):387--397, 2012.

\bibitem{Hyvonen2004}
N.~Hyv{\"o}nen.
\newblock Complete electrode model of electrical impedance tomography:
  Approximation properties and characterization of inclusions.
\newblock {\em SIAM Journal on Applied Mathematics}, 64(3):902--931, 2004.

\bibitem{Isaacson2006}
D.~Isaacson, J.L. Mueller, J.C. Newell, and S.~Siltanen.
\newblock Imaging cardiac activity by the {D}-bar method for electrical
  impedance tomography.
\newblock {\em Physiological Measurement}, 27:S43--S50, 2006.

\bibitem{article}
Diederik Kingma and Jimmy Ba.
\newblock Adam: A method for stochastic optimization.
\newblock {\em International Conference on Learning Representations}, 12 2014.

\bibitem{knudsen2009regularized}
Kim Knudsen, Matti Lassas, Jennifer~L Mueller, and Samuli Siltanen.
\newblock Regularized d-bar method for the inverse conductivity problem.
\newblock {\em Inverse Problems and Imaging}, 35(4):599, 2009.

\bibitem{8253590}
A.~{Lucas}, M.~{Iliadis}, R.~{Molina}, and A.~K. {Katsaggelos}.
\newblock Using deep neural networks for inverse problems in imaging: Beyond
  analytical methods.
\newblock {\em IEEE Signal Processing Magazine}, 35(1):20--36, 2018.

\bibitem{lunz2018adversarial}
Sebastian Lunz, Ozan {\"O}ktem, and Carola-Bibiane Sch{\"o}nlieb.
\newblock Adversarial regularizers in inverse problems.
\newblock In {\em Advances in Neural Information Processing Systems}, pages
  8507--8516, 2018.

\bibitem{malone2014stroke}
Emma Malone, Markus Jehl, Simon Arridge, Timo Betcke, and David Holder.
\newblock Stroke type differentiation using spectrally constrained
  multifrequency eit: evaluation of feasibility in a realistic head model.
\newblock {\em Physiological measurement}, 35(6):1051, 2014.

\bibitem{8103129}
M.~T. {McCann}, K.~H. {Jin}, and M.~{Unser}.
\newblock Convolutional neural networks for inverse problems in imaging: A
  review.
\newblock {\em IEEE Signal Processing Magazine}, 34(6):85--95, 2017.

\bibitem{mcdermott2018brain}
Barry McDermott, Martin O’Halloran, Emily Porter, and Adam Santorelli.
\newblock Brain haemorrhage detection using a svm classifier with electrical
  impedance tomography measurement frames.
\newblock {\em PloS one}, 13(7):e0200469, 2018.

\bibitem{mcewan2006design}
A~McEwan, A~Romsauerova, R~Yerworth, Lior Horesh, Richard Bayford, and
  D~Holder.
\newblock Design and calibration of a compact multi-frequency eit system for
  acute stroke imaging.
\newblock {\em Physiological measurement}, 27(5):S199, 2006.

\bibitem{mueller2012linear}
Jennifer~L Mueller and Samuli Siltanen.
\newblock {\em Linear and nonlinear inverse problems with practical
  applications}, volume~10.
\newblock Siam, 2012.

\bibitem{Mueller2018}
J.L. Mueller, P.A. Muller, M.M. Mellenthin, E.~DeBoer, R.~Murthy, M.~Capps,
  M.~Alsaker, R.~Deterding, and S.~Sagel.
\newblock A method of estimating regions of air trapping from electrical
  impedance tomography data.
\newblock {\em Physiological Measurement}, 39(5):05NT01, 2018.

\bibitem{cite-key}
S.~Sacco, C.~Marini, and A.~Carolei.
\newblock Medical treatment of intracerebral hemorrhage.
\newblock {\em Neurological Sciences}, 25(1):s6--s9, 2004.

\bibitem{saver2006time}
Jeffrey~L Saver.
\newblock Time is brain—quantified.
\newblock {\em Stroke}, 37(1):263--266, 2006.

\bibitem{seo2019learning}
Jin~Keun Seo, Kang~Cheol Kim, Ariungerel Jargal, Kyounghun Lee, and Bastian
  Harrach.
\newblock A learning-based method for solving ill-posed nonlinear inverse
  problems: A simulation study of lung eit.
\newblock {\em SIAM journal on Imaging Sciences}, 12(3):1275--1295, 2019.

\bibitem{shi2009experimental}
Xuetao Shi, Fusheng You, Canhua Xu, Liang Wang, Feng Fu, Ruigang Liu, and
  Xiuzhen Dong.
\newblock Experimental study on early detection of acute cerebral ischemic
  stroke using electrical impedance tomography method.
\newblock In {\em World Congress on Medical Physics and Biomedical Engineering,
  September 7-12, 2009, Munich, Germany}, pages 510--513. Springer, 2009.

\bibitem{Somersalo1992}
Erkki Somersalo, Margaret Cheney, and David Isaacson.
\newblock Existence and uniqueness for electrode models for electric current
  computed tomography.
\newblock {\em SIAM Journal on Applied Mathematics}, 52(4):1023--1040, 1992.

\bibitem{sylvester1987}
John Sylvester and Gunther Uhlmann.
\newblock A global uniqueness theorem for an inverse boundary value problem.
\newblock {\em Annals of mathematics}, pages 153--169, 1987.

\bibitem{zhou2015comparison}
Zhou Zhou, Gustavo~Sato dos Santos, Thomas Dowrick, James Avery, Zhaolin Sun,
  Hui Xu, and David~S Holder.
\newblock Comparison of total variation algorithms for electrical impedance
  tomography.
\newblock {\em Physiological measurement}, 36(6):1193, 2015.

\end{thebibliography}

\end{document}